\documentclass[twocolumn]{aastex6}
\usepackage{graphicx}
\usepackage{epstopdf}
\usepackage{amssymb,amsfonts,amsmath,amstext,amsgen,amsopn,amsxtra,indentfirst,times,subfigure,natbib}
\begin{document}
\title{\texttt{THOR}: A New and Flexible Global Circulation Model to Explore Planetary Atmospheres}
\author{Jo\~{a}o M. Mendon\c{c}a\altaffilmark{1}}
\author{Simon L. Grimm\altaffilmark{1,2}}
\author{Luc Grosheintz \altaffilmark{1}}
\author{Kevin Heng\altaffilmark{1}}

\altaffiltext{1}{University of Bern, Center for Space and Habitability, Sidlerstrasse 5, CH-3012, Bern, Switzerland.  Email: joao.mendonca@csh.unibe.ch; kevin.heng@csh.unibe.ch}
\altaffiltext{2}{University of Z\"{u}rich, Institute for Computational Science, Winterthurerstrasse 190, CH-8057, Z\"{u}rich, Switzerland.}

\begin{abstract}
We have designed and developed, from scratch, a global circulation model named \texttt{THOR} that solves the three-dimensional non-hydrostatic Euler equations. Our general approach lifts the commonly used assumptions of a shallow atmosphere and hydrostatic equilibrium. We solve the ``pole problem'' (where converging meridians on a sphere lead to increasingly smaller time steps near the poles) by implementing an icosahedral grid. Irregularities in the grid, which lead to grid imprinting, are smoothed using the ``spring dynamics'' technique. We validate our implementation of spring dynamics by examining calculations of the divergence and gradient of test functions. To prevent the computational time step from being bottlenecked by having to resolve sound waves, we implement a split-explicit method together with a horizontally explicit and vertically implicit integration. We validate our global circulation model by reproducing the Earth and also the hot Jupiter-like benchmark tests. \texttt{THOR} was designed to run on Graphics Processing Units (GPUs), which allows for physics modules (radiative transfer, clouds, chemistry) to be added in the future, and is part of the open-source Exoclimes Simulation Platform (ESP; www.exoclime.org).\end{abstract}


\section{Introduction}
\label{sec:intro}
The observational exploration of planetary atmospheres has shown that planets present a large diversity of climates and atmospheric circulations. This diversity raises the need to build versatile numerical tools capable of interpreting the observational data and help unveil the main mechanisms driving the atmospheric climate and dynamics. These tools have to be self-consistent and based on theory that does not compromise the accuracy of the results under particular planetary conditions. For this purpose, our goal is to develop the first robust Global Circulation Model (GCM)\footnote{Also termed ``general circulation models''.} capable of simulating a vast range of planetary conditions, which will work as a virtual ``planetary lab''.

Global Circulation Models (GCMs) solve the complex physical and dynamical equations that include a representation of the evolution of the resolved fluid flow and various idealizations for radiative transfer, dry or moist convection and for heat and momentum's turbulent surface fluxes. These models are powerful tools to simulate self-consistently, for example, the dynamical heat transport in the atmosphere and represent 3D temperature maps of the atmosphere essential to interpret observational data. They have been important to the study of the atmospheric circulation and climate of Earth, of the Solar System planets and more recently of extrasolar planets. In general, these planetary atmospheric simulations have been explored by changing the general parameters that characterize the planets (e.g. rotation rate, distance to the star, planetary radius, atmospheric mass and composition) and adapting the physics package to deal with a wider range of atmospheric conditions (sometimes radically different from Earth). Some of the most successful simulations of the Solar System planetary atmospheres are:  Venus (\citealt{2010Lebonnois} and \citealt{2016Mendonca}), Mars (\citealt{1999Forget}), Titan (\citealt{1995Hourdin}),  and the giant gas planets like Jupiter, Saturn, Neptune and Uranus (\citealt{1998Dowling}, \citealt{2004Yamazaki} and \citealt{2009Schneider}). Good reviews on recent advances on Earth GCMs can be found in \cite{2007Randall} and \cite{2013Dowling}. The dynamical part of these atmospheric models usually remains the same, however, that the core has been modified in \cite{2010Lebonnois} and \cite{2016Mendonca} to include, for example, variations of the specific heat with temperature for the Venus atmosphere. This numerical exploration of planetary atmospheres also expands to the extrasolar planets, as for example, the works on tidally locked hot Jupiters (e.g., \citealt{2002Showman}; \citealt{2009Showman}; \citealt{2011Heng}; \citealt{2015Kataria}). These models obtained a robust equatorial jet feature in the simulations, which seems to be a consistent phenomenon with the observational shift of the maximum flux in the secondary eclipse seen in  e.g., \cite{2009Knutson}. A good review on the models that have been used for extrasolar planets can be found in \cite{2015Heng}. The majority of the 3D atmospheric models for planetary studies used  as their basis dynamical cores that were developed to do Earth climate studies, with the exception of a model called EPIC (\citealt{1998Dowling}). EPIC model is the only 3D climate model that we are aware of, that has been developed from ground-up with the main goal of exploring different planet atmospheres.

Despite some success in planetary studies, the vast diversity of planetary characteristics observed raises questions about the flexibility of current atmospheric models to represent accurately the atmospheric physics of those planets. Planetary climate models adapted from Earth climate studies have usually included approximations that are Earth-centric. Some of the most commonly-used approximations are the shallow atmosphere and the hydrostatic approximation (described in section \ref{sec:eqs_thor}). In the new platform that we present in this work, we want to avoid making any ad hoc assumption that can compromise the physics of the problem. The amount of detailed information about extrasolar planetary atmospheres is still very limited and not enough to validate 3D climate models. Due to these limitations the modeling of extrasolar planets has to be formulated from first principles. The perfect simulation would be based on universal physical schemes with no artificial forcing that could reproduce all the observational data or help in preparing the observations (e.g., by making predictions). Those ideal ``planetary labs'' still do not exist and we are still in the important process of assessing the robustness of our models for different atmospheric conditions. A careful exploration of the parameter space needed for this process can teach us important lessons on how the atmospheric circulation and climate work. In addition, it continues to be poorly understood how new atmospheric circulations are driven for a large range of  astronomical and planetary bulk parameters. Numerically, we also need to improve our knowledge on the balance between physical and numerical sources and sinks of quantities such as angular momentum and total energy for a large diversity of planetary conditions, which can be crucial to assess the accuracy and robustness of the simulations.

The dynamical core is the part of the atmospheric models that solves the resolved dynamical fluid equations including thermodynamics and mass conservation. The new and flexible dynamical model \texttt{THOR} is part of the Exoclimes Simulation Platform (ESP), and the code is intended to be freely open-source (see more information in www.exoclime.org). The ESP is divided into two cores, the core that solves the physics equations such as radiative transfer and convection,  and the core that solves the fluid equations. The physics core used in this work is very simplified, and more sophisticated schemes are being developed to be included in the physics core. Currently, two physical schemes are being implemented: \texttt{HELIOS} and \texttt{VULCAN}. \texttt{HELIOS} is an efficient and flexible radiative transfer code suitable for 3D climate models (\citealt{2015Malik}). This code will represent the radiative emission/absorption/scattering in the atmosphere due to gas molecules and clouds in the ESP simulations. The scheme uses as input k-distribution tables, which are produced by the open-source program called \texttt{HELIOS-K} (\citealt{2015Grimm}).  Other physics modules such as \texttt{VULCAN} will represent the atmospheric chemistry, including thermochemistry and photochemistry (\citealt{2015Tsai}). In addition to the dynamical core presented here, based on a horizontally explicit and vertically implicit (HE-VI) type time integrator, we are also working on a fully explicit Riemann solver (\citealt{2016Grosheintz}).

The dynamical core \texttt{THOR} was planned to be flexible, simple to use, and suitable for parallel computations which will allow us to run efficient high-resolution simulations necessary to resolve important instabilities and to improve numerical accuracy. The main goal when developing \texttt{THOR} was to provide a solid dynamical core capable of simulating and helping us to have a better understanding on the atmospheric dynamics and climate of a large diversity of planets. 

In this work, we present the first version of the 3D atmospheric model \texttt{THOR}.  In the next section, we will start describing the spherical grid used to discretize the model domain. In this section, we also describe the necessary modifications to the icosahedral to improve the numerical accuracy of the divergence and gradient operators. In section \ref{sec:eqs_thor}, we present the equations solved by \texttt{THOR} and the methods used for the time and space integrations. We also discuss the main approximations often used in climate models and their validity. The numerical diffusion applied to the prognostic variables (these are variables predicted by the integration of the equations) in \texttt{THOR} is presented and discussed in section \ref{sec:num_diss}. In section \ref{sec:ProfX}, we include the description of the simple physics representations for radiation, convection and boundary layer friction, which were coupled with the new dynamical core. In section \ref{sec:gpu}, we briefly describe the Graphics Processing Unit (GPU) implementation of the new code. In section \ref{sec:simu}, the new model is validated against other models for two distinct atmospheric conditions. Finally, in section \ref{sec:conclusion}, we present the general conclusions.

\section{Grid structure}
\label{sec:grid}

A simple latitude-longitude grid is associated with the convergence of the meridians at the poles that largely constrains the time-step at high latitudes to maintain the model stability (known as the ``pole problem''). In order to relax the Courant-Friedrichs-Lewy condition at high latitudes the equations in \texttt{THOR} are solved in an icosahedral grid, which is a quasi-uniform grid. This polyhedron was chosen in favour of other grids that also solves the pole problem due to its higher uniformity and isotropy. The icosahedral grid was first applied to numerical atmospheric models in the 1960s by \cite{1968Sadourny} and \cite{1968Williamson}. In the beginning, these methods were not actively pursued due to the rise of other more popular methods at that time, such as spectral models. The spectral models, in general, are numerically more accurate than grid point models. However, at high resolutions they become computationally more inefficient than the grid point models. Another drawback is the poor representation of large gradients or discontinuities in the spectral models, which can result in spurious waves (this is also known as ``spectral ringing''). For these reasons models based on icosahedral grids have become recently more popular (e.g, \citealt{2002Majewski}, \citealt{2004Tomita} and \citealt{2015Zangl}), and it is the option chosen to be the core of \texttt{THOR}. In the next two subsections we explain how the icosahedral grid is built and modified to improve numerical accuracy and how the main divergence and gradient are defined in the control volumes. A good review on horizontal grids for global circulation models can be found in \cite{2012Staniforth}.

\subsection{Standard icosahedral}

\begin{figure}[!ht]
\begin{center}
\vspace{0.1in}
\includegraphics[width=1.0\columnwidth]{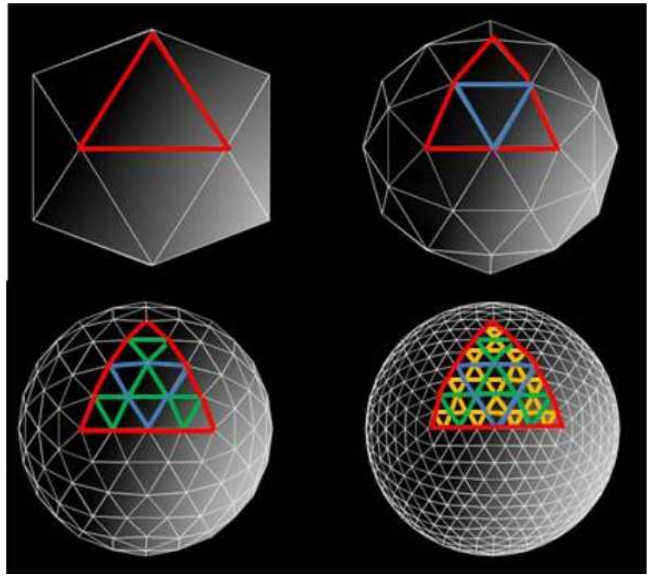}
\end{center}
%\vspace{-0.2in}
\caption{Grid refinement using a recursive method which consists in bisecting each edge and projecting the new points in the sphere.}
%\vspace{0.1in}
\label{fig:icogrid}
\end{figure}

The construction of our grid starts always from the platonic solid known as the icosahedron. This solid has twenty faces (equilateral triangles), thirty edges and twelve vertices uniformly distributed over the spherical surface. Two of the twelve vertices are placed at each pole. The icosahedral grid can also be decomposed easily in ten rhombuses from the twenty equilateral spherical triangles, which can be handy for parallel computing (e.g., \citealt{2013Heikes} and the work here). There are no uniform grids (platonic solids) with more than twelve vertices, however, there are two methods that have been used to increase the number of vertices from the initial icosahedron and keep the distribution of points quasi-uniformly distributed. The two methods are based on recursive (e.g., \citealt{1985Baumgardner} and \citealt{1995Heikes}) and non-recursive (e.g.,\citealt{1968Williamson} and \citealt{1968Sadourny}) algorithms. The non-recursive method is easier to implement than the recursive method and consists in dividing each of the twenty equilateral triangles of the original icosahedron into smaller equal sections (equilateral triangles). The intersection points from the construction of the new triangles are then projected into the spherical surface and they represent the new vertices of the grid. One of the main advantages of this method against the recursive method is the larger flexibility in the model resolution. However, the ratio of the longest and shortest distance between adjacent points is higher than in the recursive method (\citealt{2011Wang}). This disadvantage of the non-recursive method makes it less attractive, which can compromise the numerical accuracy, and for this reason we use the recursive algorithm in \texttt{THOR}. Fig. \ref{fig:icogrid} shows schematically how the recursive method works. In this method, the grid refinement is done by bisecting the edges of the triangles which creates four smaller triangles. The new vertices are then projected into the spherical surface and the procedure is repeated until the desired resolution is reached. In this work, we follow \cite{2001Tomita} to define the level of refinement in the grid, where for example g-level 5 means that the grid was refined five times. The geometrical differences between this method and the non-recursive one is the fact that when a spherical equilateral triangle is refined using the recursive method the product is three spherical isosceles triangles and one spherical equilateral triangle, and the refinement of a spherical isosceles triangle results in two isosceles triangles and two general spherical triangles (\citealt{2011Wang}). At the end of the recursive method the grid has three types of spherical triangles and the non-recursive has just one. A review on the geometric properties of the icosahedral grid can be found in \cite{2011Wang}. 

Some important quantities that help measuring the grid resolution are defined as:
\begin{align}
\label{eq:grid_qts}
&N_{vertices} = 2 + 10\times2^{2\times glevel},\\
&\overline{d} = \sqrt{\frac{2\pi}{5}}\frac{A}{2^{glevel}}\\
&\overline{\theta} = \sqrt{\frac{2\pi}{5}}\frac{1}{2^{glevel}}
\end{align}
where $N_{vertices}$ is the number of grid points in one layer, $A$ represents the radius of the planet, and $\overline{d}$ and $\overline{\theta}$ are the averaged resolution in space and angle. For example, using the Jupiter radius ($A \approx 70000$ km) as a parameter in the model and a g-level 5 refinement of the grid results in a $\overline{d} \approx 2452$ km ($\overline{\theta} \approx 2$ degrees). 

\begin{figure}[!ht]
\begin{center}
\vspace{0.1in}
\includegraphics[width=1.0\columnwidth]{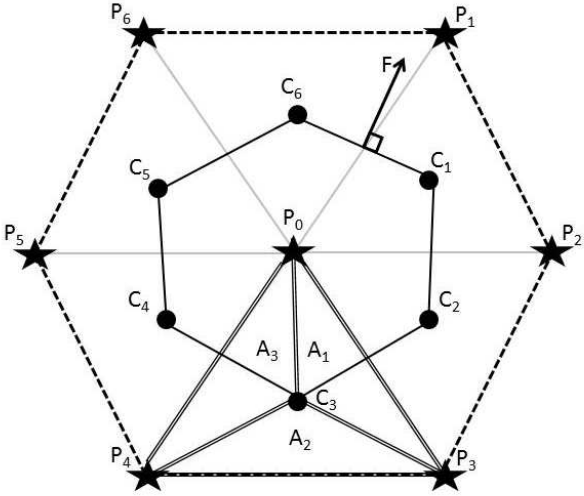}
\end{center}
%\vspace{-0.2in}
\caption{Scheme that shows how the grid is discretized for the finite-volume method. The $P$ points refer to the icosahedral vertices and the $C$ points are at the center of the triangles and they defined the control volume vertices. The areas $A_1$, $A_2$ and $A_3$ are used to interpolate the values defined in the $P$ points to the $C$ points (see Eq. \ref{eq:int_grid}). The $F$ vectors represent the numerical outward fluxes that cross the edges defined by the $C$ points.}
%\vspace{0.1in}
\label{fig:hex-fv}
\end{figure}

The icosahedral grid is covered by triangles as shown in Fig. \ref{fig:icogrid}. This grid can also be described by its dual, which is defined by connecting the centers of the triangles. The dual contains pentagons and hexagons for g-level $>$ 0. The pentagons are centered at each point of the original icosahedron and the hexagons in the other vertices. These new geometrical forms represent the control volumes in our finite-volume method as it is shown in Fig. \ref{fig:hex-fv}. We follow the method described in \cite{2001Tomita} to discretize the grid and formulate the finite-volume method. The flux at edges of the control volume needed to calculate the divergence are defined as
\begin{equation}
\label{eq:flx_div}
F_k = l_k\frac{\boldsymbol{\Phi^*}_k+\boldsymbol{\Phi^*}_{k+1}}{2}\cdot \boldsymbol{\hat{n}_k}.
\end{equation}
where $l_k$ is the geodesic arc length between the control volume vertices and $\boldsymbol{\hat{n}_k}$ is the unit outward vector normal to the geodesic arc. The asterisk in the vectors indicate that the vectors are defined at the vertices $C_k$ of the control volume (see Fig. \ref{fig:hex-fv}). The index $k$ can take a number between one and five or six, which correspond to each edge of the pentagon or hexagon respectively. To calculate the vectors at these positions we use the following equation,

\begin{equation}
\label{eq:int_grid}
\boldsymbol{\Phi}^*_k = \frac{A_2 \boldsymbol{\Phi}_0 + A_3 \boldsymbol{\Phi}_k + A_1 \boldsymbol{\Phi}_{k+1}}{A_1 + A_2 + A_3},
\end{equation}
where the vectors with no asterisk are defined at $P_k$, the cell centers. The area of the spherical triangle $A_1$ is defined by the points $P_0$, $P_k$ and $C_k$; $A_2$ by the points $P_k$, $P_{k+1}$, $C_k$; and $A_3$ by the points $P_0$, $P_{k+1}$ and $C_k$ (see Fig. \ref{fig:hex-fv}). With Eqs. \ref{eq:flx_div} and \ref{eq:int_grid} we have all the ingredients to calculate the horizontal divergence operator of a vector $\boldsymbol{\Phi}$ at point $P_0$ using the Gauss theorem, 
\begin{equation}
\label{eq:div_gauss}
\boldsymbol{\nabla_h} \cdot \boldsymbol{\Phi}_0 \simeq \frac{1}{A_c} \sum \limits_{k=1}^{6 (5)} F_k,
\end{equation}
where $A_c$ is the area of the control volume. The sum can be expanded to five or six terms depending on the number of sides of the control volume that can be a pentagon (five) or an hexagon (six). The horizontal gradient operator of a scalar $s$ is calculated using
\begin{equation}
\label{eq:grad}
\boldsymbol{\nabla_h} s_0 \simeq \frac{1}{A_c} \sum \limits_{k=1}^{6 (5)} l_k \frac{s^*_k + s^*_{k+1}}{2} \boldsymbol{\hat{n}_k} - \frac{s_0}{A_c}\sum \limits_{k=1}^{6 (5)} l_k \boldsymbol{\hat{n}_k}.
\end{equation}
The second term in Eq. \ref{eq:grad} is a correction term due to the spherical space curvature. Without this term, a gradient of a constant field would not be zero.

The vertical discretization is based on layered vertical columns that are centered at points $P$ and bounded by the control volume edges. In the numerical scheme it is often necessary to estimate physical quantities at the interfaces between two consecutive layers or to estimate the vertical momentum at the center of the layer (the vertical momentum is updated at the interfaces). To solve these problems we use an interpolation method. The model has implemented a Lagrangian polynomial interpolation algorithm based on:
\begin{align}
\Phi(z_h) &\approx \sum^k_{j=0} \Phi_j L_j(z_h), \\
L_j(z) &= \prod_{\substack{0\leq i \leq k \\ i \neq j}}\frac{z_h - z_i}{z_j - z_i},
\end{align}
where $z_i$ is the altitude at the cell center and $z_h$ at the interface. $\Phi(z_h)$ is the variable $\Phi$ interpolated to an altitude $z_h$. In the simulations explored later we use $k=1$, which is a linear interpolation, and we did not find it necessary to use higher order interpolations.   

The accuracy of divergence and gradient operators from Eqs. \ref{eq:div_gauss} and \ref{eq:grad}, when applied to the standard icosahedral described above, is low and constrains drastically the performance of the atmospheric model (see Fig. \ref{fig:MAPMN}). In the section below, we describe the modifications applied to the grid in order to improve the accuracy of the operators.

\subsection{Modified icosahedral grid}
\label{subsec:modicogr}
Numerical noise from the space integration can be expected to reveal the underlying grid structure in the numerical solution, which is associated with a numerical problem called grid imprinting. The icosahedral grid in this section is modified to increase the numerical accuracy of the divergence and gradient operators. The poor accuracy is related with the non-uniform distribution of the triangular areas. As we described above, the division of the triangles during the grid refinement results in triangles with different areas, and in \cite{2001Tomita} it is argued that regions with the largest area gradient correspond to regions with higher grid noise. The larger distortions in the grid are located adjacent to the geodesics from the initial icosahedron. These distortions can also be measured by the ratio of the longest and shortest distance between adjacent points as described in the previous section, and it is larger in regions of larger grid distortion. Despite the recursive method being smoother than the non-recursive method, we still apply a method called ``spring dynamics'' to the grid (\citealt{2001Tomita}) to smooth it. After this method is applied, we do a second step that consists of moving the new icosahedral vertices to the centroid of the control volumes, which allows the model to reach an effective accuracy close to the second-order. Note that the order of accuracy is related to the rate of convergence of the numerical solution to the exact solution, which means that the error in a second-order scheme decays quadratically with the resolution.

As pointed above, the first step to improve the quality of the grid and smooth the grid distortions is to apply ``spring dynamics'' (\citealt{2001Tomita}). In this method all the grid points are connected to their nearest neighbors by springs. The grid modified is the icosahedral grid obtained by recursive refinements and with a radius equal to one (nondimensional radius from the original icosahedron). In the ``spring dynamics'' method, the net force applied to each grid point due to the attached springs is represented by
\begin{equation} 
\label{eq:spring_force}
\frac{d\boldsymbol{v}}{dt} = \sum^{6}_{i=1} k (d_i - \overline{d}_{spr})\boldsymbol{e_i} - \alpha \boldsymbol{v}.
\end{equation}
In this equation, $\alpha$ is the friction coefficient, $k$ is the spring constant, $d$ is the arc length between the central point and one of the neighbor points which in the case shown in Eq. \ref{eq:spring_force} is between $P_0$ and $P_i$. The values of $i$ vary between 1 and 6 because the position of the center of the twelve initial pentagons remains fixed. $\overline{d}_{spr}$ is the natural spring length, and $\boldsymbol{v}$ is the velocity of the unit mass grid points. The latter is defined simply by how much the points move:
\begin{equation}
\label{eq:spr_vel} 
\boldsymbol{v} = \frac{d\boldsymbol{P_0}}{dt}
\end{equation}
The natural spring length $\overline{d_{spr}}$ can be written as
\begin{equation}
\label{eq:dbar} 
\overline{d}_{spr} = \beta\frac{2\pi A}{10\times2^{glevel-1}}.
\end{equation}
The parameter $\beta$ is a tuning parameter and it was set to 1.15 in our work as suggested by \cite{2001Tomita}. The friction term allows the system to converge smoothly to an equilibrium avoiding possible instabilities created by the sudden move of the grid points. The parameter $\alpha$ that controls the friction is a tunable parameter and we set this value to 1 for simplicity.  The dynamical system is integrated with a time step of 0.01 until the net force in each point becomes very small. The simulation is stopped following the condition that the points do not move more than $10^{-5}$ is satisfied in every point. Note that there is no need to associate the variables in this scheme to any units due to the simplicity and the geometric goal of the method. After the grid distortions were smoothed with the ``spring dynamics'' scheme, the grid points are moved to the centroid of the control volume to increase the accuracy of the finite-volume method (\citealt{2001Tomita}). In this step, we first correct the position of the vertices ($C_k$) of the control volumes by moving them to the centroid of the triangles. Next, we keep the vertices fixed and move the center ($P_i$) of the control volume to the its new centroid. The new positions are found using the following equation:
\begin{equation}
 \boldsymbol{r_c} = \frac{1}{2\pi}\sum^{n}_{i=1} \frac{\boldsymbol{r_i}\times \boldsymbol{r_{i+1}}}{|\boldsymbol{r_i}\times \boldsymbol{r_{i+1}}|}\tan^{-1}(|\boldsymbol{r_i}\cdot \boldsymbol{r_{i+1}}|/|\boldsymbol{r_i}\times \boldsymbol{r_{i+1}}|).
\end{equation}
In this equation $\boldsymbol{r_c}$ is the new centroid, $n$ can be three, five or six depending if it is the centroid of a triangle, pentagon or hexagon, and $\boldsymbol{r}$ is the point position. 

To validate the accuracy of the operators with a new modified grid, we used the test functions proposed by \cite{1995Heikes}:
\begin{align}
&\alpha(\lambda, \varphi) = \sin(\lambda),\\
&\beta(\lambda, \varphi) = \cos(m\lambda) \cos^4(n\varphi),\\\nonumber
&\boldsymbol{v} = \alpha\boldsymbol{\nabla}\beta \\\label{eq:div_test}
&\hspace{0.5cm}= -m\frac{\cos^4(n\varphi)}{\cos(\varphi)}\sin(\varphi)\sin(m\lambda)\boldsymbol{i} \\ \nonumber& \ \ \ \ - 4n\cos^3(n\varphi)\sin(n\varphi)\sin(\lambda)\cos(m\lambda)\boldsymbol{j}.
\end{align}
The quantities that we will test are $\nabla\beta$ and $\nabla\cdot \boldsymbol{v}$ for the solutions when $m=1$ and $n=1$, and when $m=3$ and $n=3$. These tests challenge the accuracy of the operators across the sphere and the accuracy can be quantified because the exact solution is known. To quantify the error in the operators, we define the identities:
\begin{equation}
l_2 = I[(\phi(\lambda,\varphi)-\phi_{exact}(\lambda,\varphi))^2]^{1/2},
\end{equation}
\begin{equation}
l_{\infty} = max|\phi-\phi_{exact}|
\end{equation}
where the function $\phi$ can be $\nabla\cdot \boldsymbol{v}$ or $\nabla\beta$ and the function $I$ represents global average. 

\begin{figure}
\centering
\subfigure[Standard grid]{\label{fig:mapm1-n1}\includegraphics[width=0.5\textwidth]{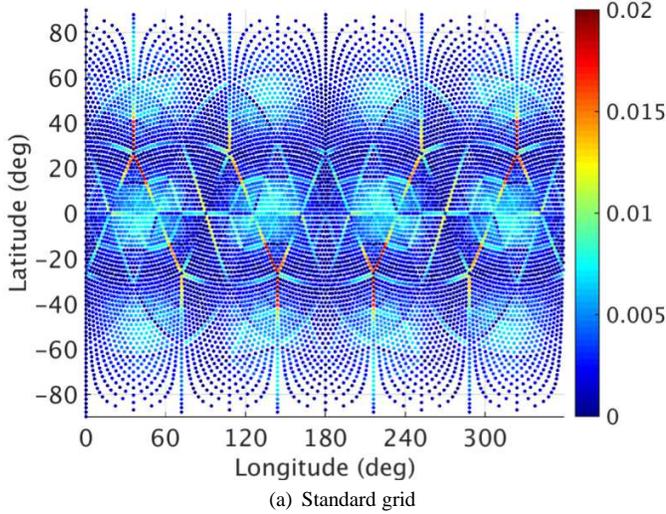}}
\subfigure[Modified grid]{\label{fig:mapm3-n3}\includegraphics[width=0.5\textwidth]{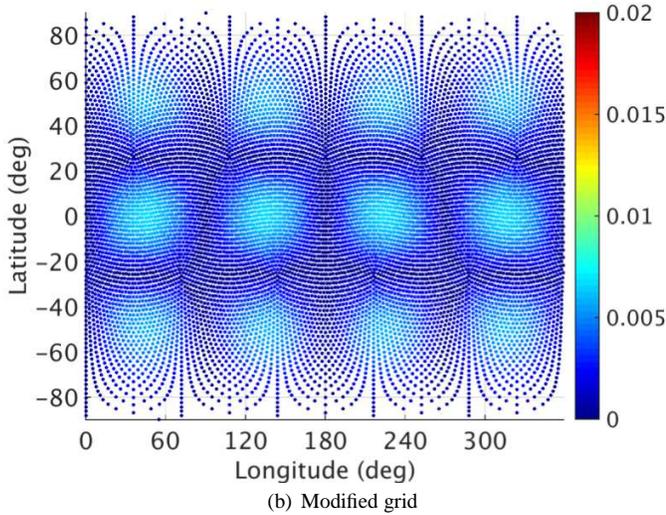}}
\caption{Absolute error for $\nabla\cdot\boldsymbol{v}$ defined in Eq. \ref{eq:div_test} when $m=1$ and $n=1$. In map (b) we apply the ``spring dynamics'' method and move the centre of the control volumes to the centroid.}
\label{fig:MAPMN}
\end{figure}

As a first step in the error analysis, we compute the absolute error of $\nabla\cdot\boldsymbol{v}$ for the solution when $m=1$ and $n=1$. Fig. \ref{fig:MAPMN} shows the distribution of the errors in the modified and unmodified grid. As we mentioned before, the largest errors are located in the regions with the largest grid distortions, which are located adjacent to the geodesic that represent the initial icosahedron. When the ``spring dynamics'' method is applied and center of the control volumes moved to the centroids, the grid noise is efficiently removed. The largest amplitudes in the modified grid case are then related to the maxima of the function $\nabla\cdot\boldsymbol{v}$. 

\begin{figure}
\centering
\subfigure[m = 1 and n = 1]{\label{fig:m1-n1}\includegraphics[width=0.5\textwidth]{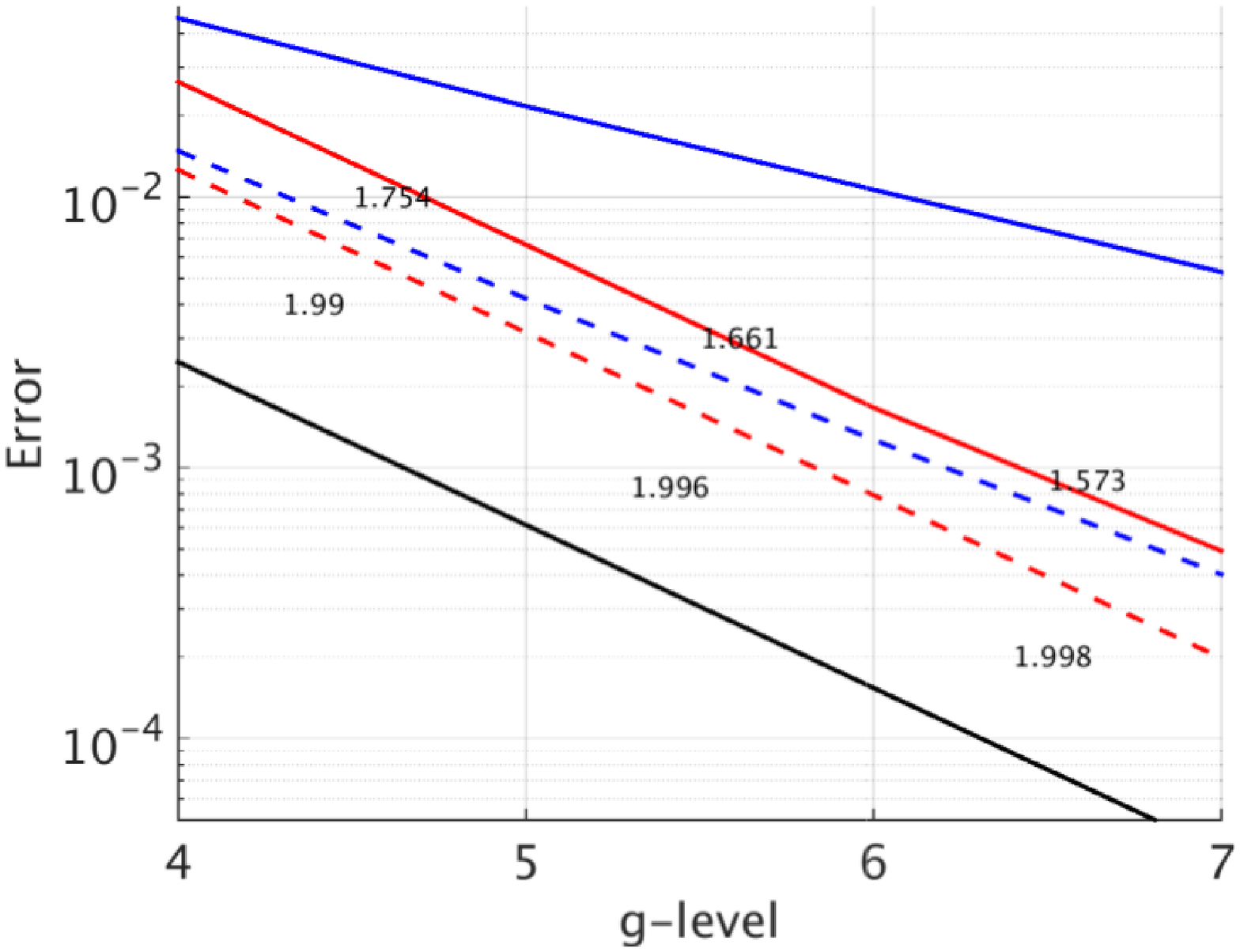}}
\subfigure[m = 3 and n = 3]{\label{fig:m3-n3}\includegraphics[width=0.5\textwidth]{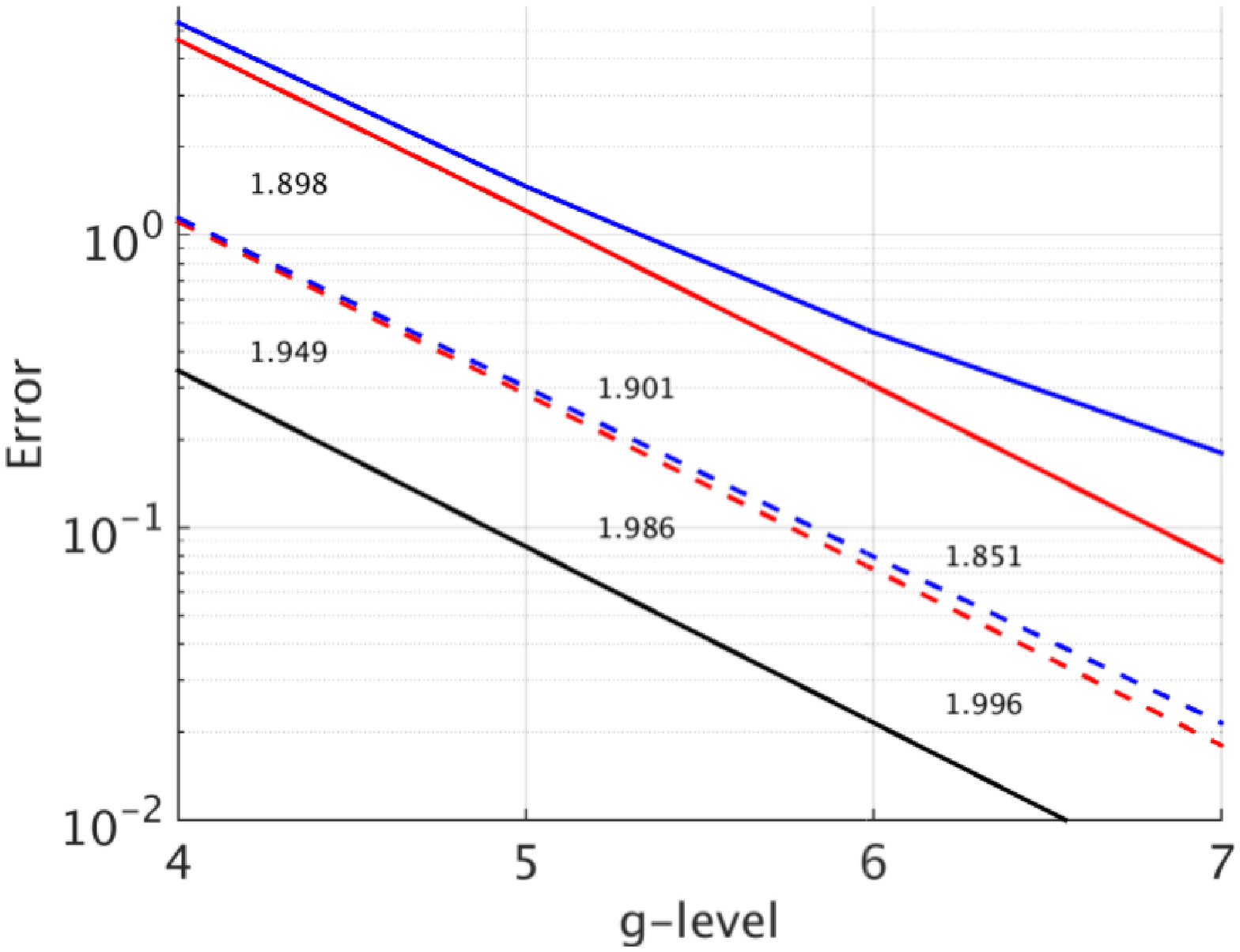}}
\caption{Error convergence plots for $\nabla\cdot\boldsymbol{v}$. Figure (a) shows the results obtained when setting the values of  $m$ and $n$ of the test functions to 1, and to 3 in figure (b). The red and blue solid lines represent the values of $l_{\infty}$ and the dashed lines $l_2$. The red lines were obtained with the modified grid and the blue lines with the unmodified grid. The black solid line is a reference line with a slope of an exact second order convergence. The numbers indicate the slopes of $l_2$ (dashed lines) for the two grids.}
\label{fig:MN}
\end{figure}

\begin{figure}
\centering
\subfigure[m = 1 and n = 1]{\label{fig:m1-n1}\includegraphics[width=0.5\textwidth]{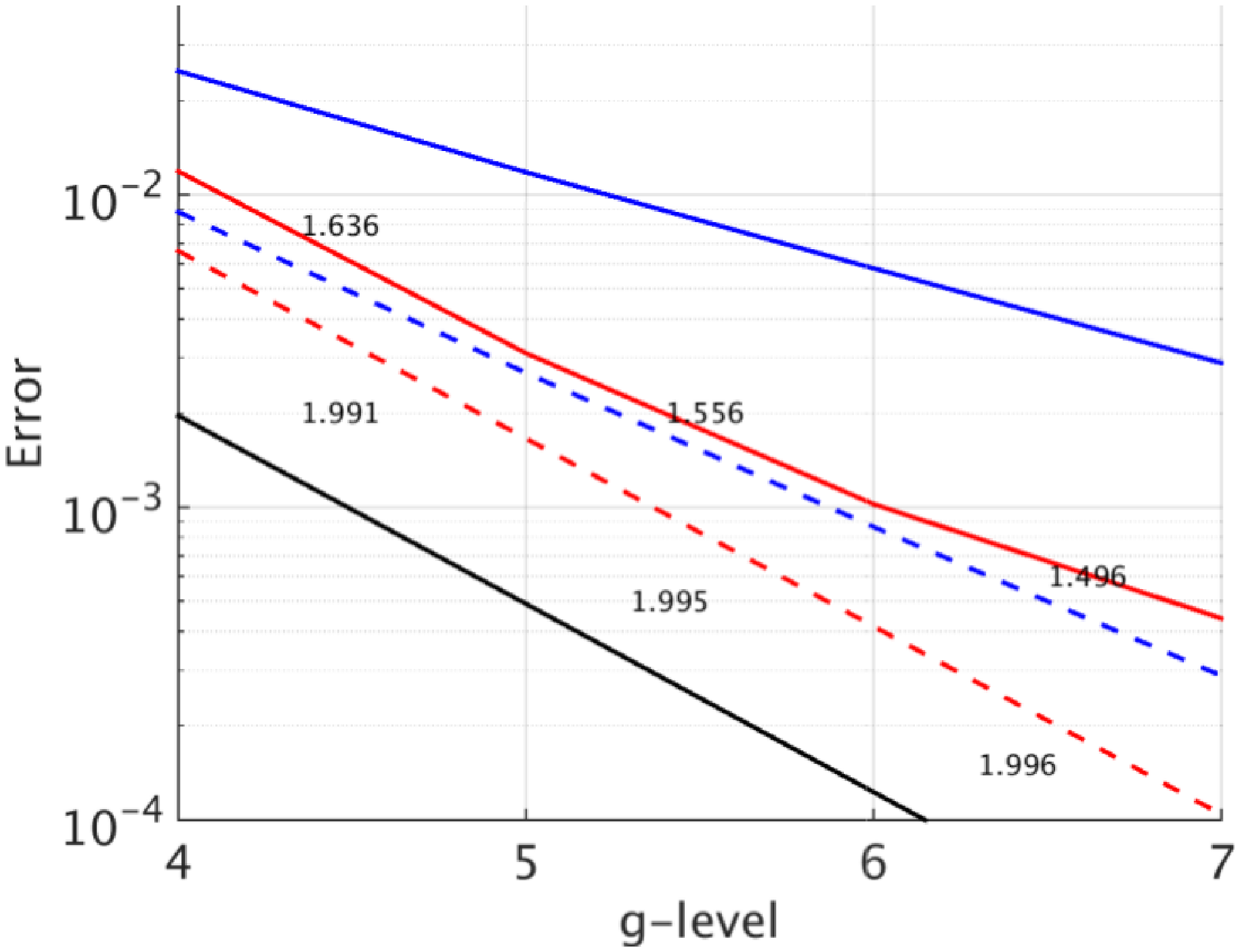}}
\subfigure[m = 3 and n = 3]{\label{fig:m3-n3}\includegraphics[width=0.5\textwidth]{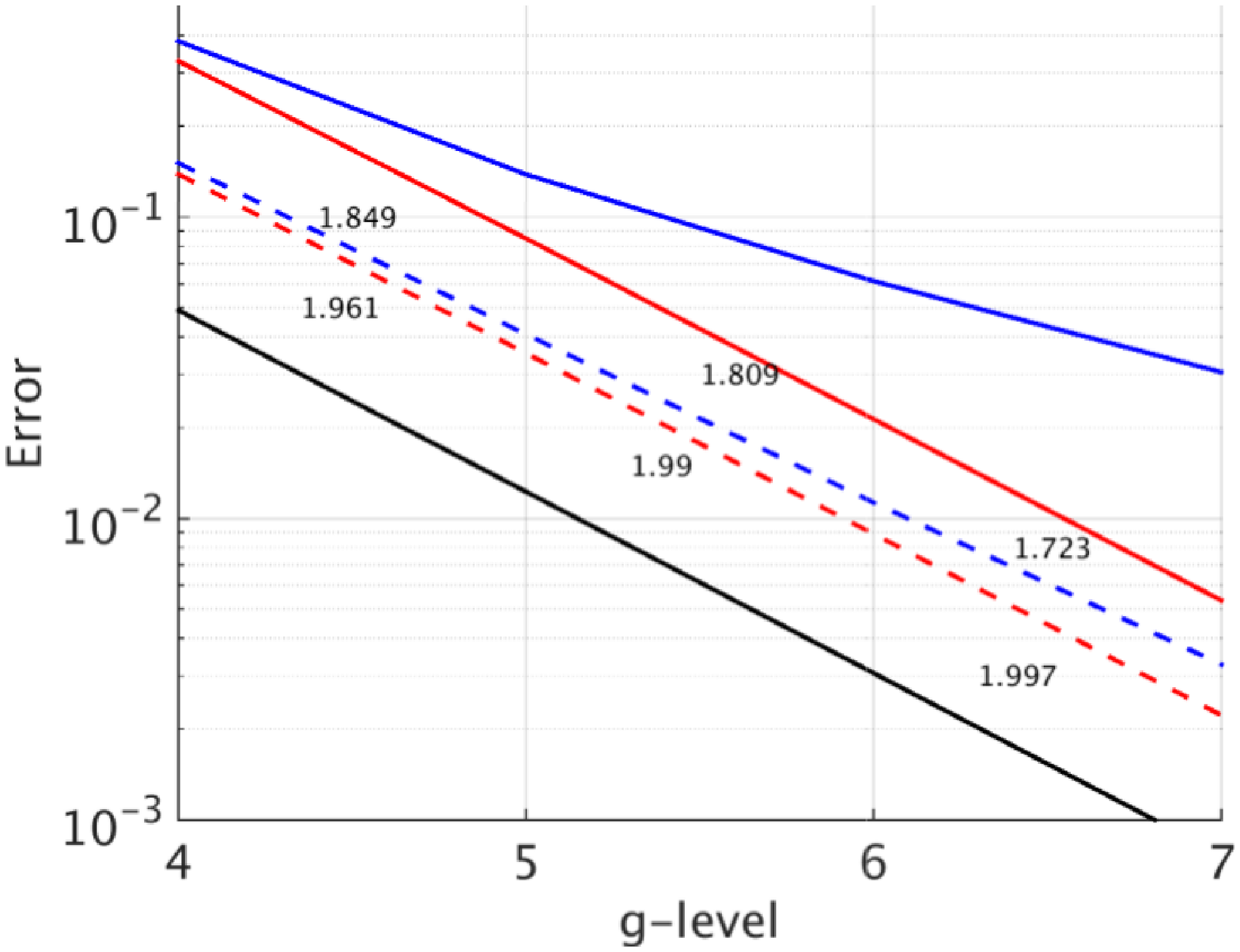}}
\caption{Error convergence plots for $\nabla\beta$. Figure (a) shows the results obtained when setting the values of  $m$ and $n$ of the test functions to 1, and to 3 in figure (b). The red and blue solid lines represent the values of $l_{\infty}$ and the dashed lines $l_2$. The red lines were obtained with the modified grid and the blue lines with the unmodified grid. The black solid line is a reference line with a slope of an exact second order convergence. The numbers indicate the slopes of $l_2$ (dashed lines) for the two grids.}
\label{fig:MN2}
\end{figure}

Now that we have gained more intuition on the location of the largest errors we explore the accuracy of the divergence and gradient operators using $l_2$ and $l_{\infty}$. Our goal when we modified the icosahedral grid was to reach an order of accuracy close to the second-order, which is the theoretical maximum order for the finite-volume scheme with a linear spacial reconstruction (\citealt{BookLeveque}). $l_2$ represents how on averaged the root mean square error is converging, and in the case of $l_{\infty}$ it is associated to the convergence of the point with the least accuracy. Figs. \ref{fig:MN} and \ref{fig:MN2} show the convergence plots. In Fig. \ref{fig:MN}, we analyze the accuracy of the model when computing $\nabla\cdot\boldsymbol{v}$ for different grid resolutions. For both test functions the grid modification clearly improved the accuracy (Figs. \ref{fig:MN} and \ref{fig:MN2}). This is also reflected in improved $l_2$ convergence rates where we reach second order for the modified grid. The $l_\infty$ case is similar to $l_2$, however, in the solution with $m=1$ and $n=1$, the last slope is slightly smaller than the $l_2$ case. The analysis of the results in Fig. \ref{fig:MN2}, for the quantity $\nabla\beta$ is very similar to Fig. \ref{fig:MN}. 

In summary, the modifications to the grid improve the accuracy of the divergence and gradient operators. Despite in some cases the convergence of $l_{\infty}$ being slower than the globally-averaged case, an accuracy close to second order is reached in all grid points. Later in section \ref{sec:Earth}, we also verify that in a long-term simulation the grid noise is not detected in the model results, and have no impact in the flow simulated.  

There are different ways of storing the physical quantities in the grid. In our current model we use the Arakawa-A grid, where all the physical variables are defined at the center of the control volume. This structure has been tested in models that use icosahedral grids, such as \cite{1996Stuhne} and \cite{2001Tomita}. This decision is associated with the set of dynamical equations that we solve in \texttt{THOR}  (presented in the next section) and the numerical scheme described above. Our strategy was to use a simple and efficient solver. Other options have been proposed in the literature for icosahedral grids, such as the $ZM-$grid in \cite{2002Ringler}. In the $ZM-$grid the velocity components and the mass are defined at different positions (staggered grid). This grid allows instabilities related to the fact that the velocity has higher resolution than the mass to grow (these instabilities are also known by computational modes). However, if the computational modes are efficiently filtered, the $ZM-$grid has advantages against all the other grids in the literature, such as a better representation of the geostrophic adjustment (\citealt{2002Ringler}). We have to use higher horizontal resolution in \texttt{THOR} to reach the same accuracy than models that use $ZM-$grids.

\section{Dynamical equations and approximations}
\label{sec:eqs_thor}

\texttt{THOR} represents resolved fluid dynamical phenomena in the atmosphere. The new model solves the 3D non-hydrostatic compressible Euler equations on a rotating sphere. Our goal was to reduce as much as possible the number of approximations usually applied to the Euler equations in Earth climate models that would compromise the flexibility of the model. These approximations simplify the equations, neglecting some physical terms, and are used mainly to improve the performance of the models without large compromises in accuracy under Earth-like conditions. However, if the accuracy of the approximations vary for different planetary atmospheres it can compromise the physics of the problem. One should be careful when adapting a climate model that has been used to do, for example, Earth climate studies since the approximations included in the model can be very Earth-centric. Some of the most often used approximations that we $avoid$ are:
\begin{itemize}
\item Hydrostatic approximation: The vertical pressure gradient balances the gravitational force. This approximation implies that the vertical scale of the atmospheric motions are smaller than the horizontal scales so that $H/L \ll 1$, where $H$ is the scale height and $L$ the characteristic horizontal scale. The hydrostatic assumption eliminates the vertical propagation of acoustic waves because the acoustic terms are neglected. If the model is able to resolve important atmospheric phenomena with horizontal scales smaller than the scale height (e.g., cumulus convection or fine-scale gravity waves), then this approximation can compromise the accuracy of the results.   
\item Shallow atmosphere approximation: The spherical components of the fluid equations that depend on $1/r$ are changed to the constant $1/A$, where $A$ is the mean planetary radius. This approximation simplifies the equations and it is justified if the atmosphere explored is much thinner than the distance from the center of the planet to the region simulated. However, this approximation has extra costs as pointed out in \cite{1966Phillips}. In order to conserve axial angular momentum some metric terms have to be neglected and also the Coriolis terms involved in the horizontal components of planet's angular velocity. These approximations are known as the ``traditional approximations''.  
\item Anelastic approximation: The density in the divergence term of the mass continuity equation is replaced by a reference density that only depends on altitude. The original formulation for the anelastic approximation was suggested by \cite{1962Ogura}. The idea of this approximation was to neglect the elasticity of the fluid and consequently filter sound waves. This approximation allows larger steps in the time integration than in the integration of the full compressible model without assuming hydrostatic balance. This approximation assumes that the physical thermodynamic quantities have just small variations from the basic state.  
\end{itemize}

The approximations described above are not used in this work, however, they have been implemented in the model and the user can switch them on and off to explore their impact on the results. Despite not using any of these approximations we still make some assumptions about the atmospheres being simulated, such as the atmosphere not being ionized and the effective gravity being constant and radial. In the future we will develop modules which will complement and improve the physical representation of these two cases. The first case becomes important in highly irradiated/very hot atmospheres and requires a 3D magnetohydrodynamic treatment of the equations (e.g., \citealt{2014Rogers}). The assumption for constant effective gravity breaks down if we are exploring deep atmospheres. The radial constant gravity  assumption can be easily fixed in the equations,  however, it is a good approximation of the two test atmospheres explored in this work. In fast rotating planets, such as Jupiter, where the centrifugal forces ($-\boldsymbol{\Omega}\times(\boldsymbol{\Omega}\times\boldsymbol{r})$) are not negligible compared to the Newtonian gravitational forces ($-\nabla \Phi_N$), spherical geopotential surfaces become a poor approximation to represent the geopotential surfaces (\citealt{2005White}). In this case, the apparent gravity is defined as:
\begin{equation}
\boldsymbol{g} \equiv -\nabla \Phi = -\nabla (\Phi_N - \frac{\Omega^2s^2}{2}),
\end{equation}
where $s$ is the perpendicular distance to the rotation axis. Since the apparent gravity is a dominant force, it is convenient to express the fluid equations in a  spheroidal coordinate system as suggested in e.g., \cite{2014Staniforth}.

The equations in \texttt{THOR} are written in the flux-form, which helps conserving our prognostic variable in the model domain. A good performance of a dynamical core requires that the timescale of spurious numerical source/sinks of conservative quantities are longer than the physical source/sink timescales. The conservation properties of quantities that have the longest physical source/sink timescales are the ones that require more attention (\citealt{2008Thuburn}). The three main equations that describe the flow in a dry atmosphere in \texttt{THOR} are:
\begin{align} \label{eq:mass}
&\frac{\partial \rho}{\partial t} + \boldsymbol{\nabla} \cdot (\rho \boldsymbol{v}) = 0,\\\label{eq:mom}
&\frac{\partial \rho \boldsymbol{v}}{\partial t} + \boldsymbol{\nabla} \cdot (\rho \boldsymbol{v} \otimes \boldsymbol{v})  = -  \boldsymbol{\nabla}p - \rho g\boldsymbol{\hat{r}} -2\rho\boldsymbol{\Omega}\times\boldsymbol{v} ,\\\label{eq:entropy}
&\frac{\partial \rho\theta}{\partial t} + \boldsymbol{\nabla} \cdot (\rho\theta\boldsymbol{v}) = 0
\end{align}

The first equation represents the mass conservation and $\rho$ is the atmospheric density and $\boldsymbol{v}$ the velocity. Mass conservation is one of the most fundamental conservation properties in climate models. Dry mass is a robust invariant in the model and it should be conserved independently of any diffusion or friction processes acting in the atmosphere. Any spurious perturbations in the mass distribution will have an impact in the pressure field, which affects, for example, the flow motion. 

Eq. \ref{eq:mom} represents Newton's second law of a fluid motion. In this equation, $p$ is the atmospheric pressure, $\Omega$ is the planetary rotation rate, $g$ is the gravity, $\boldsymbol{\hat{r}}$ is the unit vector in the radial direction and $\otimes$ is the tensor product. This equation is essential in the representation of the atmospheric flow. The momentum in the atmosphere is not a robust invariant unlike the mass described above, because the momentum is transferred to smaller scales or it can be dissipated at the boundary layer. This transfer of kinetic energy is described in section \ref{sec:num_diss}. The second term on the left side of the equation is a non-linear term called the advection term. On the right we have the balance of three forces: the first one is produced by the pressure gradient, the second is the gravitational force and the last one is related to the planet's rotation and is called the Coriolis force.  

Eq. \ref{eq:entropy} is the thermodynamic equation and it is formulated in the flux-form for entropy. For this equation, we followed the work from \cite{1990Ooyama}, \cite{2008Skamarock} and \cite{2012Ullrich}.  Entropy is closely linked with potential temperature $\theta$, where the specific entropy $S$ can be defined by:
\begin{equation}
dS = C_p d(\ln\theta)
\end{equation}
where $C_p$ is the specific heat at constant pressure. From our set of equations, it is essential to recover the pressure that is used to solve Eq. \ref{eq:mom}. The pressure is calculated from:
\begin{equation}
\label{eq:pressure}
p = p_{ref}\Big(\frac{R_d(\rho\theta)}{p_{ref}}\Big)^{\frac{C_p}{C_v}}.
\end{equation}
In the equation above the variable $p_{ref}$ refers to a reference pressure level and for the simulations tested in section \ref{sec:simu} $p_{ref}$ was set to 1 bar. The variable $R_d$ is the gas constant for dry air and $C_v$ is the specific heat at constant volume. 

This thermodynamic equation does not include explicitly the conversion between mechanical energy into thermal energy. In a real atmosphere there is transfer of kinetic energy from large to smaller scales, and at some point this transfer of energy crosses the truncation limit of the model. This energy is dynamically transported and eventually dissipated by frictional heating (molecular scale), and it is reintroduced into the resolved scales by internal energy. Part of this cycle happens at unresolved scales in the model. In \cite{2012Rauscher} the dissipative kinetic energy is reintroduced directly into the energy cycle as internal energy. Another possible solution to close this cycle is to solve the total energy equation in the flux-form  (\citealt{2004Tomita}) instead of the entropy equation used in \texttt{THOR}. These solutions would close the cycle since they are conserving the total energy, however, it can lead to local spurious conversions between available and unavailable potential energy, which would result in a poor performance of the model. There are also numerical schemes that inject back some of the dissipative energy into the resolved kinetic energy using a stochastic pattern generator in order to improve the atmospheric circulation at scales close to the truncation limit (\citealt{2005Shutts}). However, it is not clear how can the rate of energy backscatter be generalized for planetary applications. We are still investigating what are the best options to represent the missing physics that can be important in atmospheres that have large transfer of kinetic energy to unresolved scales. We have implemented and tested, in order to keep energy balance in the model, a scheme proposed by \cite{2009Williamson}, which globally uniform adjusts the internal energy at all grid points. Numerical inaccuracies in the total energy conservation can lead to drifts in the mean circulation (e.g., \citealt{1998Boville}) and this scheme helps to alleviate the problem. However, due to its low impact in the results of the simulations explored in this work, the results presented later were obtained without any energy fixer scheme.

To complement the main set of equations we use the assumption that the atmosphere can be treated as an ideal gas, which is a good approximation for the atmospheric conditions explored in this work:
\begin{equation}
p = \rho R_d T,
\end{equation}
where $T$ is the atmospheric temperature. A more general equation of state can be included in \texttt{THOR}, which implies two modifications: one, on how the pressure is estimated after the entropy equation is integrated and two, on how the pressure equation is used to do the implicit vertical integration of the momentum.

The 3D divergence and gradient operators are split as: 
\begin{align}
& \boldsymbol{\nabla}\Phi =\boldsymbol{\nabla}_h\Phi + \boldsymbol{\hat{r}}\frac{d\Phi}{dr}, \\
& \boldsymbol{\nabla\cdot\Phi} = \boldsymbol{\nabla}_h\cdot\boldsymbol{\Phi_h}+\frac{1}{r^2}\frac{d}{dr}(r^2\Phi_r), 
\end{align}
where $r$ is the distance from the planet center, $\boldsymbol{\hat{r}}$ is the unit radial vector, and  $\boldsymbol{\nabla}_h$ and $\boldsymbol{\nabla}_h\cdot$ are the spherical gradient and divergence operators defined in Eqs. \ref{eq:div_gauss} and \ref{eq:grad}. In the second equation the variables $\boldsymbol{\Phi_h}$ and $\Phi_r$ are the horizonal (parallel to the spherical surface) and radial projections of the vector $\boldsymbol{\Phi}$. These two quantities are calculated in the model using:
\begin{equation}
\label{eq:proj_h}
 \boldsymbol{\Phi_h} =  \boldsymbol{\Phi} -  \boldsymbol{\Phi}\cdot  \boldsymbol{\hat{r}},
\end{equation}
\begin{equation}
\label{eq:proj_r}
 \Phi_r =  \boldsymbol{\Phi}\cdot  \boldsymbol{\hat{r}},
\end{equation}
where $\boldsymbol{\hat{r}}$ is the unit radial vector.

\texttt{THOR} solves the full Euler equations that allow the development of, for example, fast acoustic waves that constrains the time step needed to satisfy the Courant-Friedrichs-Lewy (CFL) condition.  To overcome this restriction we use a split-explicit method from  \cite{2002Wicker} and \cite{2008Skamarock}, coupled with a HE-VI method (Horizontal Explicit - Vertical Implicit,\citealt{2004Tomita}). The split-explicit method involves splitting the time integration of the Euler equation in short explicit steps for the terms associated with acoustic modes and large steps for the rest. The time-splitting method is simple to implement and improves the efficiency of the integration allowing the CFL condition to be obeyed in the horizontal integration. In the vertical direction, we have typically much higher resolution than in the horizontal one, and the fast vertically propagating waves are handled with an implicit scheme. Other techniques have been explored to make non-hydrostatic models sound-proof. One example is the semi-implicit method where the horizontal and vertical propagation of sound waves is treated implicitly (\citealt{2012Durran}). However, this method requires the expensive numerical  solution of a 3D elliptic equation. More recently,  \cite{2014Dubos} extended the work from  \cite{2009Arakawa} and derived the equations for a sound-proof semi-hydrostatic model from Hamilton's principle of least action. The new set of equations retains the vertical acceleration term which provides accurate model results at non-hydrostatic scales and filters efficiently the acoustic waves from the atmospheric flow due to the constrain to the air parcels to stay close to their hydrostatic position. But as pointed out in  \cite{2014Dubos}, it is still not clear if the integration of the new equations is more efficient than the integration of the full Euler equations with the method used in our work ( \citealt{2008Skamarock}). Note that in this work we are not interested in resolving the sound waves that are in general energetically very low and have almost no impact in the atmospheric circulation and climate. To resolve correctly the sound waves with \texttt{THOR}, the time step of the time integration has to be reduced until the CFL condition for these fast waves is satisfied in any direction.

Eqs. \ref{eq:mass}, \ref{eq:mom} and \ref{eq:entropy} are divided in terms related with fast and slow physical processes. This splitting, as we said above, is done to improve the performance of the time integration. The strategy is to integrate the different terms with different numerical methods: the fast terms, related with the compressible terms that generate fast acoustic waves, are integrated with a simple and fast scheme and using a short time step; and the slow terms with a more accurate scheme and longer time step. Fig. \ref{fig:time_scheme} shows schematically how the time integration method works. The time-splitting method used in this work is similar to the ones described in \cite{2002Wicker}, \cite{2004Tomita} and \cite{2008Skamarock}. For the integration of the slow terms, we use a time integration scheme proposed by \cite{2002Wicker}, which is a predictor-corrector scheme with second-order accuracy:
\begin{align}\nonumber
& \Phi^{(1)} = \Phi^{[t]} + \frac{\Delta t}{3}R(\Phi^{[t]}) \\
& \Phi^{(2)} = \Phi^{[t]} + \frac{\Delta t}{2}R(\Phi^{(1)}) \\\nonumber
& \Phi^{[t + \Delta t]} = \Phi^{[t]} +\Delta t R(\Phi^{(2)})
\end{align}
where $\Delta t$ is the large time step, $\Phi$ is a prognostic variable and $R$ the equation terms that estimate the rate of $\Phi$. In the inner loop, we use a forward-backward scheme with a short time step. The number of inner loops $n$, and also the time step of the short steps $\Delta \tau$ change for each large time step $\Delta t$ (\citealt{2007Klemp}). In the first large time step the inner loop is just done once with a time step equal to $\Delta t/3$, in the second large step the inner loop runs $n/2$ times with a time step of $\Delta t/n$, and in the last large step the inner loop runs for $n$ times with a time step of $\Delta t/n$. A large value for $n$ improves the stability of the model but it becomes very time expensive. In the simulations explored later we set the value of $n$ to 6. To improve the accuracy of the split-explicit method the equations are integrated using a perturbation formulation. During the inner loop the deviations ($\Phi^\star$) from the large time step ($\Phi^{[t]}$), are defined as
\begin{equation}
\Phi^\star = \Phi - \Phi^{[t]}.
\end{equation}
Below we write the equations solved where $\tau$ is the time step of the inner loop and $t$ for the large time step:
\begin{equation}
\begin{split}
\frac{{\rho^*} ^{[\tau + \Delta \tau]} -  {\rho^*} ^{[\tau]}}{\Delta \tau}  + \boldsymbol{\nabla}_h \cdot (\rho{\boldsymbol v_h)^*}^{[\tau + \Delta \tau]}\hspace{1cm} \\
+ \frac{1}{r^2}\frac{\partial}{\partial r} {(\rho v_r)^*} ^{[\tau + \Delta \tau]}r^2 = - \boldsymbol{\nabla}_h \cdot (\rho{\boldsymbol{v}_h})^{[t]} \\
- \frac{1}{r^2}\frac{\partial}{\partial r} (\rho{v_r}) ^{[t]}r^2,
\label{eq:dens2}
\end{split}
\end{equation}
\begin{equation}
\begin{split}
\frac{(\rho{\boldsymbol{v}_h)^*} ^{[\tau + \Delta \tau]} - (\rho{\boldsymbol{v}_h)^*} ^{[\tau]}}{\Delta \tau} + \boldsymbol{\nabla}_h {p^*}^{[\tau]} = - \boldsymbol{\nabla}_h {p}^{[t]}  \\ 
-\boldsymbol{\cal A}_h^{[t]} - \boldsymbol{\cal C}_h^{[t]} ,
\label{eq:mom2}
\end{split}
\end{equation}
\begin{equation}
\begin{split}
\frac{(\rho{\boldsymbol{v}_r)^*} ^{[\tau + \Delta \tau]} - (\rho{\boldsymbol{v}_r)^*} ^{{[\tau]}}}{\Delta\tau} +  \frac{\partial}{\partial r}{p^*}^{[\tau + \Delta\tau]} \\
+ {\rho^\star}^{[\tau + \Delta\tau]}g = -\frac{\partial}{\partial r}{p}^{[t]} - \rho^{[t]}g - {\cal A}_r^{[t]} -  {\cal C}_r^{[t]},
\label{eq:momr2}
\end{split}
\end{equation}
\begin{equation}
\begin{split}
\frac{{(\rho\theta)} ^{[\tau + \Delta \tau]} -  {(\rho\theta)} ^{[\tau]}}{\Delta \tau}  + \boldsymbol{\nabla}_h \cdot {\theta^{[t]} (\rho\boldsymbol{v}_h})^{[\tau + \Delta \tau]} + \\
\frac{\partial}{\partial r} \theta^{[t]} (\rho{v_r})^{[\tau + \Delta \tau]} = 0.
\label{eq:pt2}
\end{split}
\end{equation}
The slow terms are written in the right side of the equations. The variables $\boldsymbol{\cal A}$ and $\boldsymbol{\cal C}$ are the advection and Coriolis terms. These two terms are calculated using the non-projected velocities, and then projected parallel to the spherical surface and in the radial direction using the Eqs. (\ref{eq:proj_h}) and (\ref{eq:proj_r}). We do not calculate the deviation of $\theta$ from the large step directly from Eq. (\ref{eq:pt2}) to avoid negative values which would compromise the conversion to pressure values needed for Eqs. (\ref{eq:mom2}) and (\ref{eq:momr2}). Instead we do:
\begin{equation}
\label{eq:theta_to_p}
{p^*}^{[\tau + \Delta\tau]} = p_{ref}\Big(\frac{R_d(\rho\theta)^{[\tau+\Delta\tau]}}{p_{ref}}\Big)^{\frac{C_p}{C_v}} - p^{[t]}
\end{equation}
The first equation integrated in the model is the momentum equation (\ref{eq:mom2}), which is done using an explicit method. After the horizontally projected momentum is updated we solve the implicit equation (\ref{eq:momr2}). Integrating the vertical momentum using an implicit method, we improve the performance of the model due to the instabilities created by the fast propagation of the sound waves in the typically fine vertical resolution while keeping a reasonable time step. The method used to solve Eq. (\ref{eq:momr2}) in a nonhydrostatic framework is called the HE-VI (horizontal explicit and vertical implicit) scheme and it was proposed in \cite{2002Satoh} and \cite{2004Tomita}. Eqs. (\ref{eq:dens2}) and (\ref{eq:pt2}) are arranged for ${\rho^*}^{[\tau + \Delta\tau]}$ and ${p^*}^{[\tau + \Delta\tau]}$, and the two solutions are inserted into Eq. (\ref{eq:momr2}). The new equation for the vertical momentum is:
\begin{equation}
\begin{split}
&-\frac{1}{r^2}\frac{\partial^2}{\partial r^2}r^2h^{[t]}(\rho v_r)^{\star[\tau+\Delta\tau]}+ \frac{2}{r^3}\frac{\partial}{\partial r}r^2h^[t](\rho v_r)^{\star[\tau+\Delta\tau]}\\
&-\frac{\partial}{\partial r}\tilde{g}^{[t]}(\rho v_r)^{\star[\tau+\Delta\tau]} - \frac{C_v}{R_d}\frac{g}{r^2}\frac{\partial}{\partial r} r^2(\rho v_r)^{\star[\tau+\Delta\tau]}\\
&+\frac{Cv}{R_d \Delta\tau^2}(\rho v_r)^{\star[\tau+\Delta\tau]} = \frac{Cv}{R_d}\Big[\frac{1}{\Delta\tau^2}(\rho v_r)^{\star[\tau]}\\
&- \frac{\partial}{\partial r}S_p-\frac{1}{\Delta\tau}\frac{\partial}{\partial r}p^{[\tau]} -\frac{g}{\Delta\tau}(S_\rho+\rho^{[\tau]})+\frac{S_{v_r}}{\Delta\tau},
\label{eq:vr22}
\end{split}
\end{equation}
where $h$ is the enthalpy and $\tilde{g}$ is the effective gravity defined by:
\begin{align}
h^{[t]} = C_pT^{[t]},\\
\tilde{g}^{[t]} = \frac{1}{\rho^{[t]}}\frac{\partial p^{[t]}}{\partial r}.
\end{align}
The functions $S_p$, $S_\rho$ and $S_{v_r}$ need to calculate $C_0$ are computed from:
\begin{align}
&S_p = -\frac{R_d}{C_v}\Big(\boldsymbol{\nabla}_h\cdot (h^{[t]}{(\rho \boldsymbol{v}_h)^*}^{[\tau+ \Delta\tau]})-\boldsymbol{\nabla}_h\cdot (h^{[t]}{(\rho \boldsymbol{v}_h)}^{[t]})\\\nonumber
&\hspace{0cm}-\frac{1}{r^2}\frac{\partial}{\partial r}(h^{[t]}{(\rho v_r)}^{[t]}r^2)+\boldsymbol{v}_h^{[t]}\cdot\boldsymbol{\nabla}p^{[t]}-\tilde{g}^{[t]}(\rho v_r)^{[t]}\Big),\\
&S_\rho = -\boldsymbol{\nabla}_h\cdot ({\rho \boldsymbol{v}_h)^*}^{[\tau+ \Delta\tau]} -\boldsymbol{\nabla}_h\cdot ({\rho \boldsymbol{v}_h)}^{[t]}-\frac{1}{r^2}\frac{\partial}{\partial r}(\rho v_r)^{[t]}r^2, \\
&S_{v_r} = -\frac{\partial}{\partial r}p^{[t]} - \rho^tg - A_r^{[t]} - C_r^{[t]}. 
\end{align}

After all the coefficients are calculated, we use a staggered grid to solve Eq. (\ref{eq:vr22}). This equation is discretized in a finite-difference form over intervals of $\Delta z$ (thickness of the layers), written as a tridiagonal matrix, and solved numerically using a Thomas algorithm. Using this method, the vertical momentum is updated at the model layer interfaces. This solver is applied to every vertical column in the grid. The solution requires two boundary conditions at the top and bottom of the model domain. In our current version of \texttt{THOR}, we set the vertical velocities to zero at the boundaries. As shown in \cite{2003Staniforth}, the only way to formally enforce total energy and mass conservation in simulations is applying a rigid lid in time and space at the boundaries.

After the vertical momentum equation is solved, we update the new densities from Eq. (\ref{eq:dens2}). This equation is calculated using an implicit formulation and the new updated momentum values. Finally we integrate Eq. (\ref{eq:pt2}) to compute the new updated potential temperatures.
\begin{figure}
\centering
\label{fig:time_scheme}\includegraphics[width=0.58\textwidth]{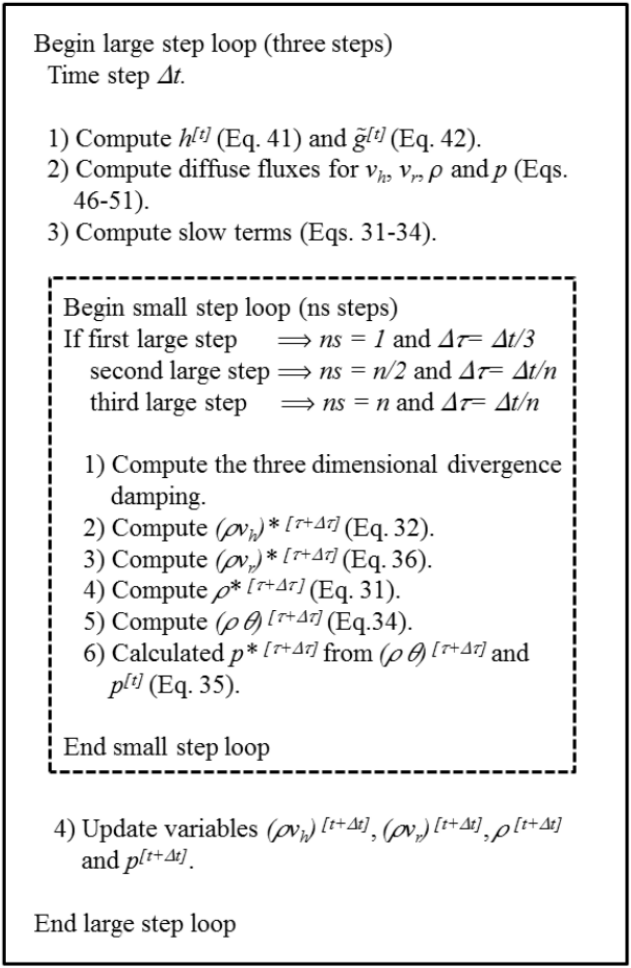}
\caption{Time integration scheme.}
\label{fig:time_scheme}
\end{figure}

\section{Numerical dissipation}
\label{sec:num_diss}
Numerical dissipation schemes are an important part of atmospheric models. These routines have an important role keeping the model numerically stable, and every model needs a representation of numerical dissipation, which can be associated with inherent dissipation from the numerical methods used or represented explicitly in the numerical scheme. These schemes are intended to alleviate the accumulation of potential enstrophy (enstrophy quantifies the intensity of the vorticity) at the smallest grid scales, which could be produced by physical downscale cascade, numerical noise from the space and time integration, misrepresentation of nonlinear interactions (aliasing effect) or poor initial conditions. Physical interpretations of the diffusion schemes are often poor, or absent, and one of the main efforts when developing these schemes is to link them to simple representations of turbulence and eddy viscosity in the sub-grid scale. 

The strength of the numerical dissipation is associated with the damping time scales of the waves present in the simulation. This quantity can have an important impact on large scale phenomena and in the case of inappropriately strong diffusion it can make fine structures disappear and affect the wave-mean flow interaction in the atmosphere, but at the same time it needs to ensure an ideal energy dissipation rate at the truncation grid scales (e.g., \citealt{2011Thrastarson}). In \cite{2011Heng}, it is shown that the calibration of the numerical horizontal dissipation can have an important impact in the estimation of the temperature and wind magnitude values under typical hot Jupiter conditions. Our current model uses an explicit dissipation representation and the strength of the diffusion is a tunable parameter, which was chosen to be large enough to ensure a stable model integration, and it does not change with latitude or longitude. Currently, we are also investigating other possibilities for the diffusion strength, such as the ones based on nonlinear formulations (\citealt{1963Smagorinsky}). Another issue that is being studied is the impact of a formulation of the numerical dissipation based on symmetric Reynold stress tensors that can guarantee the conservation of the axial angular momentum in the control volumes (see \citealt{2001Becker}). A good and comprehensive review on numerical dissipation methods in atmospheric models can be found  in \cite{2011Jablonowski}.

The order of the hyper-diffusion operator is related to the number of times $\nabla^2$ is applied to a prognostic variable. In the case of applying the operator once we have a second-order diffusion. This low-order diffusion can be associated to, for example, heat diffusion or molecular diffusion, but since these are phenomena that work at much smaller scales than the ones resolved by any global circulation model, these low order schemes should not be used. However, some models use second-order diffusion as numerical sponge schemes at the top of the model domain to effectively reduce the spurious reflections with the rigid lid at the top. In all the results explored with \texttt{THOR}, there was no indication that a sponge layer was needed, so we have not included it in the model. The use of any sponge layer in the model would also compromise the axial angular momentum conservation in the model top layers.  The need to implement higher order schemes in numerical models is related to the prediction from 2D turbulence theory, which tell us that in the infinite Reynold number limit the energy dissipation rate is zero (\citealt{1995Sadourny}). This means that higher-order schemes are more scale-selective and help to maximize the ratio of enstrophy to energy dissipation.  Increasing the scheme's order affects, however, the model's performance due to the increase of complexity in the equations by the recursive application of $\nabla^2$, and a compromise is done between performance and accuracy. Another potential problem with higher-order operators is related to the increase of the amplitudes of overshoots and undershoots with the increase of the operator order. In our work, we use fourth-order operators formulated in Cartesian coordinates that have the advantage of simplifying the equations since there is no need to deal with metric terms. A misrepresentation of these geometric terms can also lead to problems in the axial angular momentum conservation.

The extra fluxes from the explicit fourth-order hyper-diffusion applied to the prognostic variables in this work have the following form:
\begin{align} 
\label{eq:diff_rho}
F_\rho = -  \boldsymbol{\nabla}_h^2 K_d \boldsymbol{\nabla}_h^2 \rho,\hspace{4.2cm} \\\nonumber
F_{v_{hx}} = - \boldsymbol{\nabla}_h^2 K_d \boldsymbol{\nabla}_h^2 v_{hx}  - K_d\boldsymbol{\nabla}_h^2 \boldsymbol{\nabla}_h (\boldsymbol{\nabla}_h\cdot (\rho \boldsymbol{v_h}) \\\label{eq:diff_vx}+ \frac{1}{r^2}\frac{\partial}{\partial r}(\rho v_r r^2)),  \\\nonumber
F_{v_{hy}} = -  \boldsymbol{\nabla}_h^2 K_d \boldsymbol{\nabla}_h^2 v_{hy}  - K_d\boldsymbol{\nabla}_h^2 \boldsymbol{\nabla}_h (\boldsymbol{\nabla}_h\cdot (\rho \boldsymbol{v_h}) \\\label{eq:diff_vy}+ \frac{1}{r^2}\frac{\partial}{\partial r}(\rho v_r r^2)),  \\\nonumber
F_{v_{hy}} = -  \boldsymbol{\nabla}_h^2 K_d \boldsymbol{\nabla}_h^2 v_{hz}  - K_d\boldsymbol{\nabla}_h^2 \boldsymbol{\nabla}_h (\boldsymbol{\nabla}_h\cdot (\rho \boldsymbol{v_h}) \\\label{eq:diff_vz}+ \frac{1}{r^2}\frac{\partial}{\partial r}(\rho v_r r^2)),  \\\label{eq:diff_vr}
F_{v_r} = -  \boldsymbol{\nabla}_h^2 \rho K_d \boldsymbol{\nabla}_h^2 v_r,\hspace{3.8cm}\\\label{eq:diff_p}
F_{p} = -  R_d\boldsymbol{\nabla}_h^2 \rho K_d \boldsymbol{\nabla}_h^2 T.\hspace{3.6cm}
\end{align}

$F_\rho$ is the diffuse flux for density, $F_{v_{hx}}$, $F_{v_{hy}}$, $F_{v_{hz}}$  and $F_{v_r}$ are the dissipative fluxes for the momentum projections and $F_p$ is the diffuse flux for the pressure field. This last term is applied after the pressure is updated.

The second term in Eqs. (\ref{eq:diff_vx}), (\ref{eq:diff_vy}) and (\ref{eq:diff_vz}) is the 3D divergence damping, and is used in our model with the main purpose of damping the high-frequency gravity noise mainly produced by the the multiple-stage time integration explained in section \ref{sec:eqs_thor} (\citealt{1992Skamarock}). The strength of the 3D divergence damping was chosen in this work to be the same as the one used for the hyper-diffusion $K_d$. The divergence damping term is updated every short time step while the other diffuse terms are updated at the large step (\citealt{2004Tomita}).

In order to improve the accuracy of the $\boldsymbol{\nabla}_h^2$ calculation, we first compute the gradient using Eq. (\ref{eq:grad}) in the primary grid (triangles), which place the calculated vectors at the corners of the control volume avoiding an extra interpolation step. Then using these vectors we apply Eq. \ref{eq:div_gauss} to the control volume to obtain  $\boldsymbol{\nabla}_h^2$.

The value of $K_d$ once found for a set of particular planet conditions can be rescaled if the horizontal resolution or time is changed. For this rescaling method, we follow the work from \cite{2004Tomita} for a fourth-order scheme:
\begin{equation}
\label{eq:kd}
 K_d = {\cal D}\frac{\overline{d}^4}{\Delta t},
\end{equation}
 where $\Delta t$ is the time step, ${\cal D}$ is a nondimensional diffusion parameter and $\overline{d}$ is the average grid spacing defined in Eq. \ref{eq:grid_qts}. The damping time scale ($\tau_d$) can be estimated using
\begin{equation}
\label{eq:tau_d}
  \tau_d = \frac{\overline{d}^4}{2^{5}K_d}.
\end{equation}
Note that if any change is done to the space or time resolution the strength of the diffusion may have to be adapted.

\section{PROFX - Physics Core}
\label{sec:ProfX}

The physics core called PROFX contains the physical parameterizations that represent radiation, convective adjustments and the mechanical interaction between surface and atmosphere. In this work we use highly simplified physics schemes, to simplify the interpretation of the results of the complex dynamical core. The prognostics variables updated in the physical core are computed using a backward Euler step from the products of the dynamical core (\citealt{2012Ullrich}). For the boundary layer, we use the following equation:
\begin{equation}
\label{eq:im_eq_v}
  (\rho\textbf{v}_h)^{new} = \frac{(\rho\textbf{v}_h)^{dyn}}{1 + K_v(\sigma^{dyn}) \Delta t}
\end{equation}
where $(\rho\textbf{v}_h)^{new}$ is the new updated variable from the physics scheme, $\Delta t$ is the time step (the same as in the dynamical core for the large steps), $(\rho\textbf{v}_h)^{dyn}$ is the updated variable from the dynamical core and $K_v$ controls the strength of the damping. The variable $\sigma$ is defined as the ratio between the pressure layer and the surface pressure. The surface pressure was estimated extrapolating the pressures to the surface using the hydrostatic equation and keeping the temperature constant in that region. This equation forces the horizontal wind speed to zero inside the boundary layer region defined by $K_v$. 

For the radiation/convection representation, the temperature is forced using:
\begin{equation}
\label{eq:im_eq_t}
T^{new}=\frac{T^{dyn}+K_T(\sigma^{dyn})\Delta t T_{eq}}{1+K_T(\sigma^{dyn})\Delta t}
\end{equation}
In Eq. \ref{eq:im_eq_t}, $T^{new}$ and $T^{dyn}$ are the updated temperatures from the physical and dynamical cores respectively, and  $K_T$ is the Newtonian cooling function towards radiative-convective equilibrium. $T_{eq}$ is the prescribed basic state temperature field. When the temperature at a particular location is larger than $T_{eq}$, the heating rate is negative ($Q<0$) and the opposite when the temperature is cooler than $T_{eq}$. In next section the form of the functions $K_v$ and $K_T$ in the Eqs. \ref{eq:im_eq_v} and \ref{eq:im_eq_t} are defined for two different atmospheres.

\section{GPU Implementation}
\label{sec:gpu}
We are currently analysing the best strategies to boost the performance of \texttt{THOR}. Our existing program runs on a Graphics Processing Unit (GPU) environment. We integrate our simulations in a NVIDIA Tesla K20 GPU card, which has 2496 cores, a memory bandwidth of 208 GB/s, supports double precision floating point and includes error checking and correction (ECC) memory.  ECC memory comprises extra bits of memory intended to detect and fix memory errors, which is important to ensure consistency between results computed at different times.

GPUs and Central Processing Units (CPUs) have different architectures. The main advantage of using GPUs is the massively parallel architecture. In general, high performance simulations in the GPU reach better power- and cost-efficiency than in the CPU. For this reason, GPU has become an attractive solution for scientific computing applications (high performance at relatively low cost). Several Earth GCMs around the world are already being accelerated with GPU technology. Some examples are: ASUCA (\citealt{2010Shimokawabe}), GEOS-5 (\citealt{2011Putnam}) and WRF (\citealt{2008Michalakes}).   In these examples, significant speedups of the simulations due to GPU integration have been reported (tens of times compared to simulations using the conventional CPU).

The programming language adopted in this work is CUDA-C. This language is essentially C/C++ with extra commands and functions that allow, for example, to store data in the GPU memory and execute in parallel calculations in the many threads available in the GPU.  Codes in CUDA-C are programmed in both CPU and GPU, which are referred to as host and device, respectively. Our code is divided into two parts. In the first part, the model variables are declared, initialized and allocated in the host and device memories. In this part, we also construct the model’s grid, mathematical operators (e.g., divergence and gradient) and set the initial conditions of the atmosphere. At the end, we transfer the data from the host to the device memory.  In the second part, we execute the multiple kernels (functions executed in the device) that solve the dynamical and physical core’s equations, and transfer the results back to the host. Since the data transfer between the host and the device through the Peripheral Component Interconnect (PCI) express is significantly expensive, we perform all the computations of the dynamical and physical cores in the GPU. Currently, the prognostic variables are transferred back to the CPU to be stored every 1000 time steps, which is a parameter that can be easily adapted.  The kernels launched in the dynamical core execute individual threads that are grouped in 2D blocks of 256 threads. Within this compact block, the different threads synchronize and communicate via the fast on-chip shared memory. The block halo is accessed from a list of points stored in the global memory. To decompose the grid domain in sub-domains (thread blocks) we take advantage of the geometrical properties of the icosahedron, which can be divided into ten rhombuses.  To optimize the simulations, by selecting the appropriate size of the blocks, we divide the main rhombuses into smaller sub-domains (smaller rhombuses) as suggested in \cite{2002Randall}.

\section{Benchmarking}
\label{sec:simu}
In this section, we explore two different atmospheres in two very distinct regimes: Earth (``Held-Suarez'' test from \citealt{1994Held}) and a hot Jupiter (e.g., \citealt{2009Menou} and \citealt{2011Heng}). The goal of these two tests is to assess the model performance for long integrations under different forcings and analyze the main statistical dynamical quantities of the atmosphere. The ``Held-Suarez'' test from \citealt{1994Held} is often used in the Earth climate community as a benchmark test of dynamical cores, and the hot Jupiter test case has also been explored by different models in the astrophysics community (e.g., \citealt{2009Menou}, \citealt{2011Heng}, \citealt{2013Bending} and \citealt{2014Mayne}). The vast literature on these tests is essential in the validation and analysis of our methods to solve the Euler equations. These tests are not aimed at reproducing a realistic atmospheric circulation since they use very simple representations of adiabatic heating and dissipative processes. However, the simple parameterizations force the model to a climate state qualitatively similar than what is expected for each planet. It also allows an easier study of the nonlinear processes at work in the atmospheric circulation, to verify if the main atmospheric circulation drivers are well represented and check if any atmospheric phenomena are produced for unphysical reasons.

\begin{table}
\begin{center}
\caption{Planet parameters for the test cases.}
\begin{tabular}{ | l | l | l |}
\hline
 Parameters &  Earth & Hot-Jupiter \\ \hline \hline
 Planetary radius (m) & 6.371$\times10^6$ & $1.0\times10^8$ \\ \hline
 Planetary rotation rate  (s$^{-1}$) & 7.292$\times10^{-5}$ &  2.1$\times10^{-5}$ \\ \hline
 Surface gravity (m/s$^2$) & 9.8 &  8.0 \\ \hline
 Specific heat (J/K/kg) & 1004.6 & 13226.5 \\ \hline
 Gas constant (J/K/kg) & 287.04 & 3779.0 \\ \hline
Diffusion timescale (s) & 6.46$\times$10$^3$ & 941.2 \\ \hline
\end{tabular}
\end{center}
\label{tab:testcases}
\end{table} 

The simulations explored in this work were integrated over a period of 1200 Earth days. In both simulations the first 200 days were discarded from the analysis, since the data in this period is regarded as part of the spin-up phase of the model. The number of grid cells is also the same in the two experiments. In the construction of the icosahedral grid we apply five times the recursive method obtaining a grid g-level 5 which has 10,242 points quasi-uniformly distributed horizontally ($\approx 2$ degrees resolution), and for each point we use 37 vertical layers. 

The results presented in the next section use pressure as the vertical axis, which is done to better compare our results with other groups that presented the results in the same format. This step is not done during the model integration but as a post-processing step where the values are interpolated linearly to a pressure grid. 

We do not explore the case of a deep hot Jupiter test as shown in \cite{2011Heng} because the vertical domain of the physical scheme that forces the temperature is not suitable for the vertical coordinate system used in \texttt{THOR}. In \texttt{THOR} the vertical domain is defined in altitude and in \cite{2011Heng} it is defined in pressure. Note that this is a problem related with the formulation of the experiment and not related with the ability of \texttt{THOR} to simulate deep atmospheres. Using the equations presented in \cite{2011Heng} we found that at the highest layers the heating rates were unrealistically large due to the low pressures that were being obtained outside of the pressure range defined for the test. \cite{2014Mayne} uses a model with the same vertical axis as in \texttt{THOR} and they correct the heating rates for the lowest pressure to avoid the numerical problems and be able to simulate the deep atmospheric test case. We defer such work to a future paper.

\subsection{Benchmark test for Earth}
\label{sec:Earth}

We start the validation of the model by exploring the ``Held-Suarez'' test (\citealt{1994Held}). In this test the adiabatic heating and frictional forcing are parametrized to represent an atmospheric circulation qualitatively similar to Earth. The two parameterizations are based on two linear forcing schemes represented by Eqs. (\ref{eq:im_eq_v}) and (\ref{eq:im_eq_t}). 
In Eq. (\ref{eq:im_eq_v}) the strength of the frictional damping is defined by:
\begin{equation}
\label{eq:im_eq_fv}
  K_v(\sigma) = k_f \times max \Big(0.0, \frac{\sigma - \sigma_b}{1 - \sigma_b}\Big), 
\end{equation}
where $k_f^{-1} = 1$ Earth day, and $\sigma_b = 0.7$.  
The functions $T_{eq}$ and $K_T$ from the adiabatic heating function (Eq. $\ref{eq:im_eq_t}$) are given by:
\begin{align} 
T_{eq} = &max\Big\{200 K, \Big[315 K - (\Delta T)_y\sin^2\phi - \\\nonumber
&(\Delta\theta)_z \log\Big(\frac{p}{p^*}\Big)\cos^2\phi\Big]\Big(\frac{p}{p^*}\Big)^k\Big\},\\
K_T = &k_a + (k_s - k_a)\times\Big(0, \frac{\sigma - \sigma_b}{1 - \sigma_b}\Big)\cos^4\phi.
\end{align}
In the above equations, $k_a^{-1} = 40$ Earth days, $k_s^{-1} = 4$ Earth days, $(\Delta T) = 60$ K, $k = R_d/C_p$ and $\phi$ is the latitude. The structure of the function $T_{eq}$ represents a stable stratification at low latitudes which then decreases to zero polewards. This poleward decrease reduces the excitation of gravitational instabilities (\citealt{1994Held}).

This simulation started from a rest atmosphere and with the same temperature profile in every column taken from the function $T_{eq}$ at the equator in hydrostatic equilibrium. Starting the simulation from a non-isothermal atmosphere helps to break the symmetry between the two hemispheres. This problem is related to the existence of an unstable symmetric solution for the Held-Suarez experiment. A sudden breaking of the symmetry during the simulation can lead to the formation of large instabilities which can lead to unrealistic values. There are other techniques to break the symmetry, such as an isothermal atmosphere with small perturbations (\citealt{1994Held}) or initial perturbations in the vorticity field (\citealt{2011Heng}).

The time step used was 1000 seconds, a grid refinement equals to g-level 5, 37 vertical layers covering from 0 to 32 km and a diffusion timescale of $6.46\times10^3$ seconds. The simulation has 200 Earth days of spin-up phase where the data results are discarded. During this time the available potential energy stored in the atmosphere is converted to kinetic energy. This conversion is initially forced by the temperature forcing and instabilities induced by the planet's rotation. Before the end of the spin-up phase time, the atmospheric circulation has reached a statistical equilibrium state, where for example the total axial angular momentum has a statistically steady average value as a function of time.

\begin{figure}
\centering
\subfigure[Eastward wind speed]{\label{fig:u-hs}\includegraphics[width=0.445\textwidth]{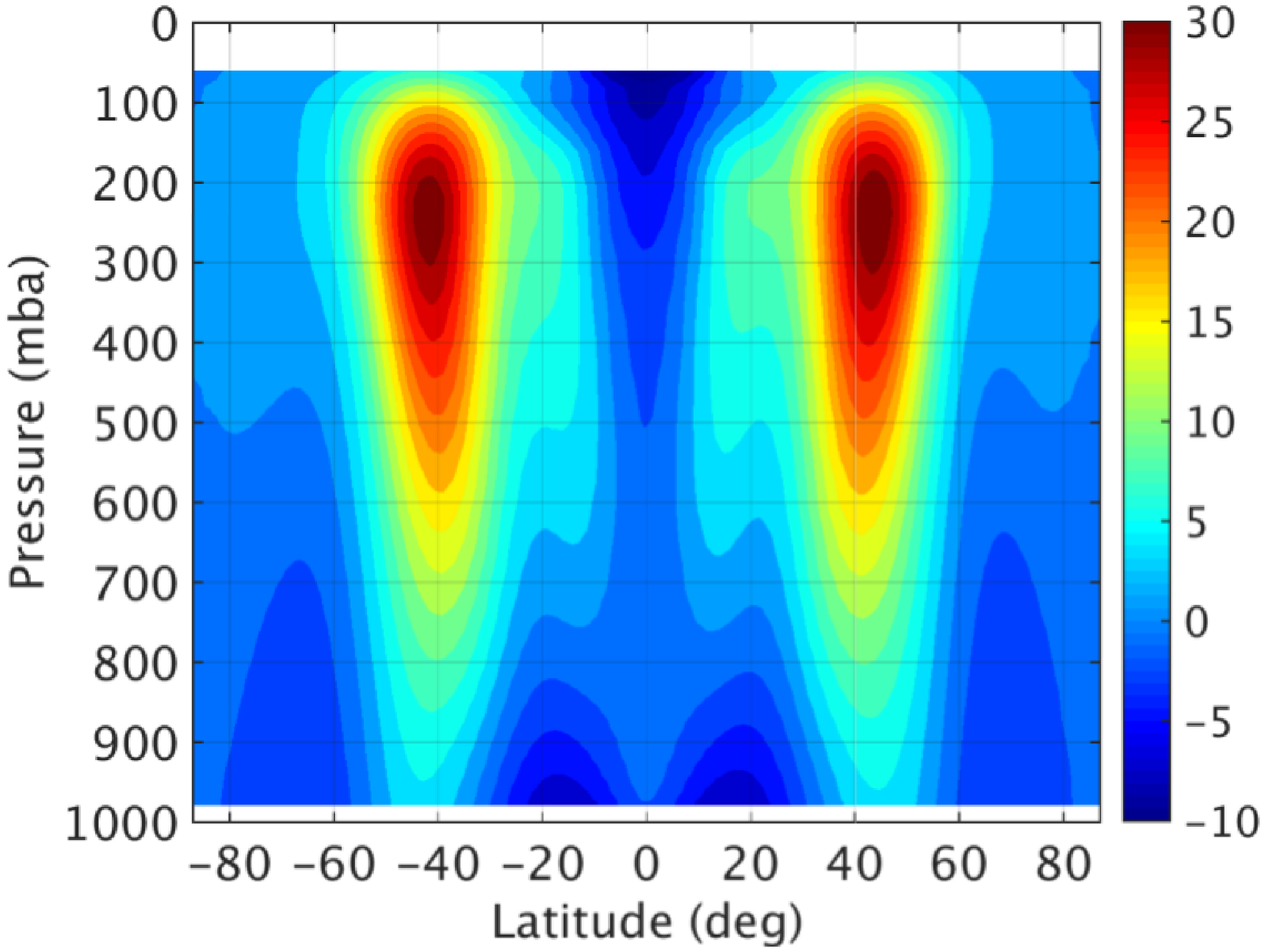}}
\subfigure[Northward wind speed]{\label{fig:v-hs}\includegraphics[width=0.445\textwidth]{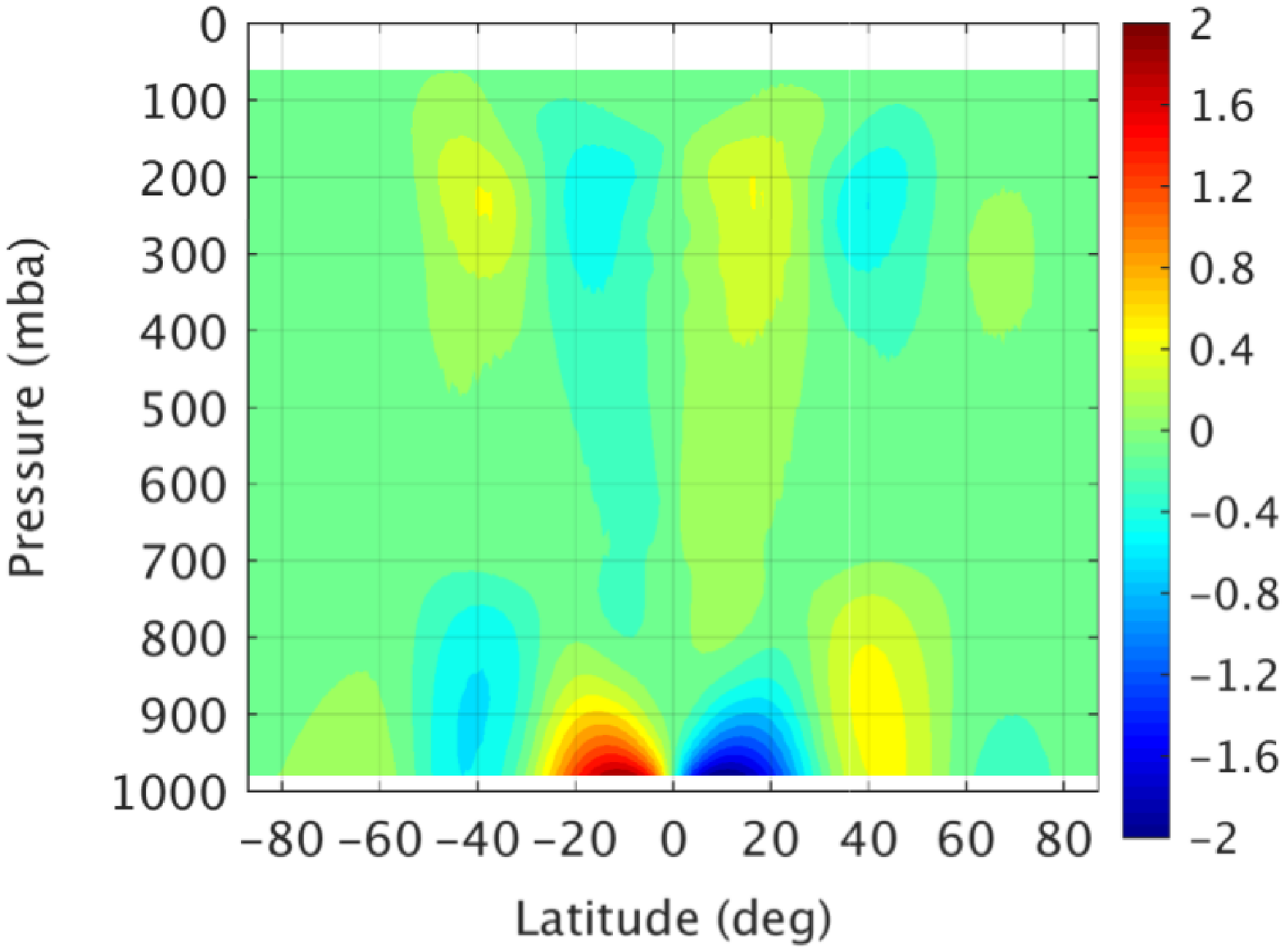}}
\subfigure[Mass stream function]{\label{fig:sf-hs}\includegraphics[width=0.445\textwidth]{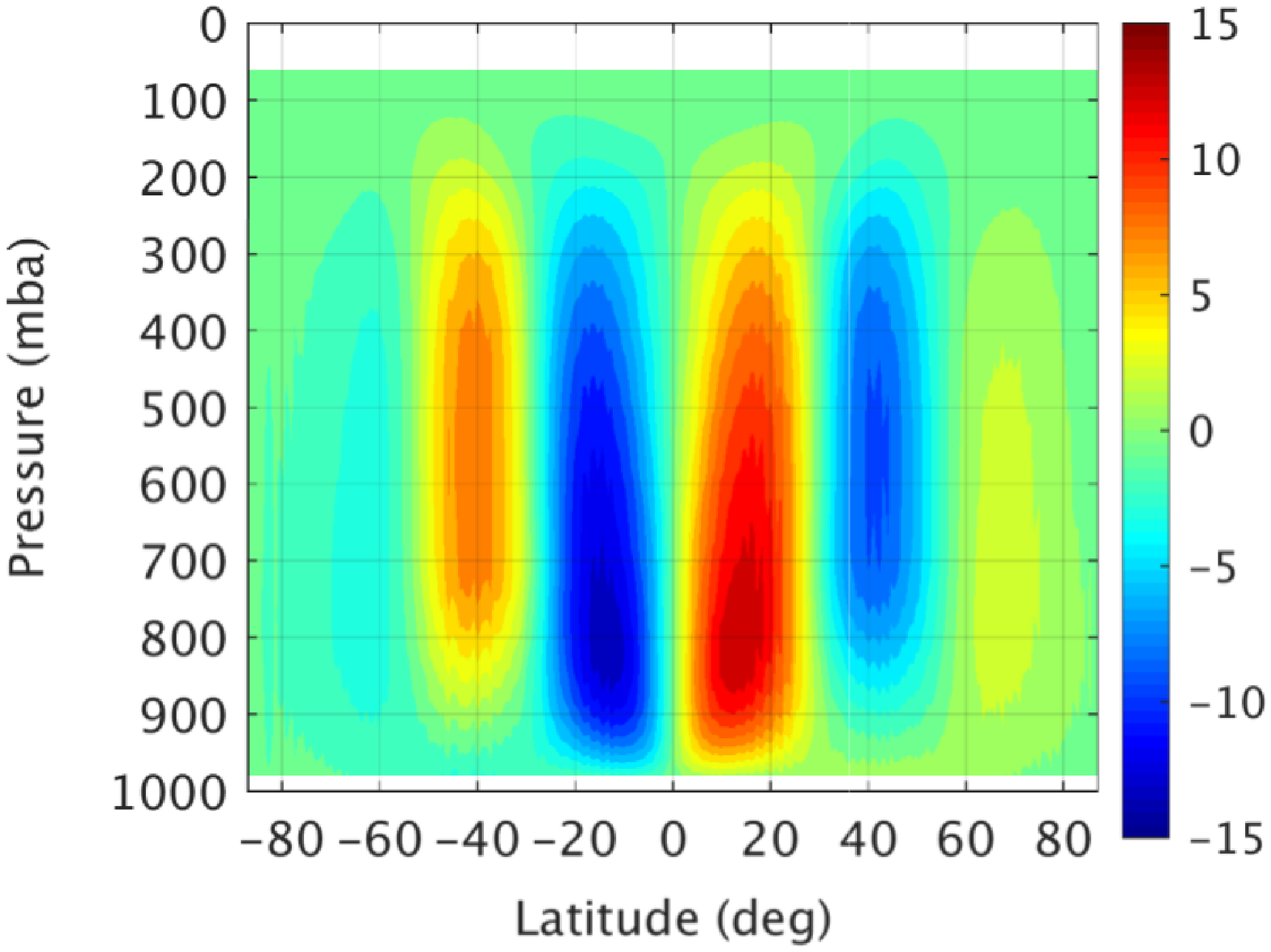}}
\caption{Final averaged zonal and meridional winds (m/s) and mass stream function ($10^{10}$ kg/s) for the Held-Suarez experiment. The values were time averaged for 1000 Earth days.}
\label{fig:HS-EXP-UV}
\end{figure}

In Figs. \ref{fig:HS-EXP-UV}, we show the zonal and meridional wind and the mass stream function for the Held-Suarez experiment. The mass streamfunction ($\Psi$) is calculated using the expression
\begin{equation}
\Psi = \frac{2\pi A \cos{\phi}}{g}\int_0^P \overline{[v]} dp.
\end{equation}
In this equation, $A$ represents the planet radius, $p$ is the pressure and $\overline{[v]}$ the zonal and time averaged meridional wind. The simulation produces one jet in each hemisphere's center at roughly 250 mbar and $45^o$ latitude consistent with the results from \cite{1994Held}. The zonal wind structure matches closely the results obtained in \cite{1994Held}. The zonal winds change sign in some regions in the atmosphere: at low latitudes in the upper atmosphere, and in the lower atmosphere it appears at low latitudes and also at latitudes higher than $60^o$. The meridional component of the wind and the mass stream function show the classical atmospheric structure in the upper atmosphere: the Hadley cells at low latitudes, followed by the Ferrel cells that extend until 60$^o$ latitude and then the polar cells (see also \citealt{2011Hengb} for a comparison). In the lower atmosphere the meridional winds change direction and become slightly stronger.

\begin{figure}
\centering
\subfigure[Temperature]{\label{fig:t-hs}\includegraphics[width=0.5\textwidth]{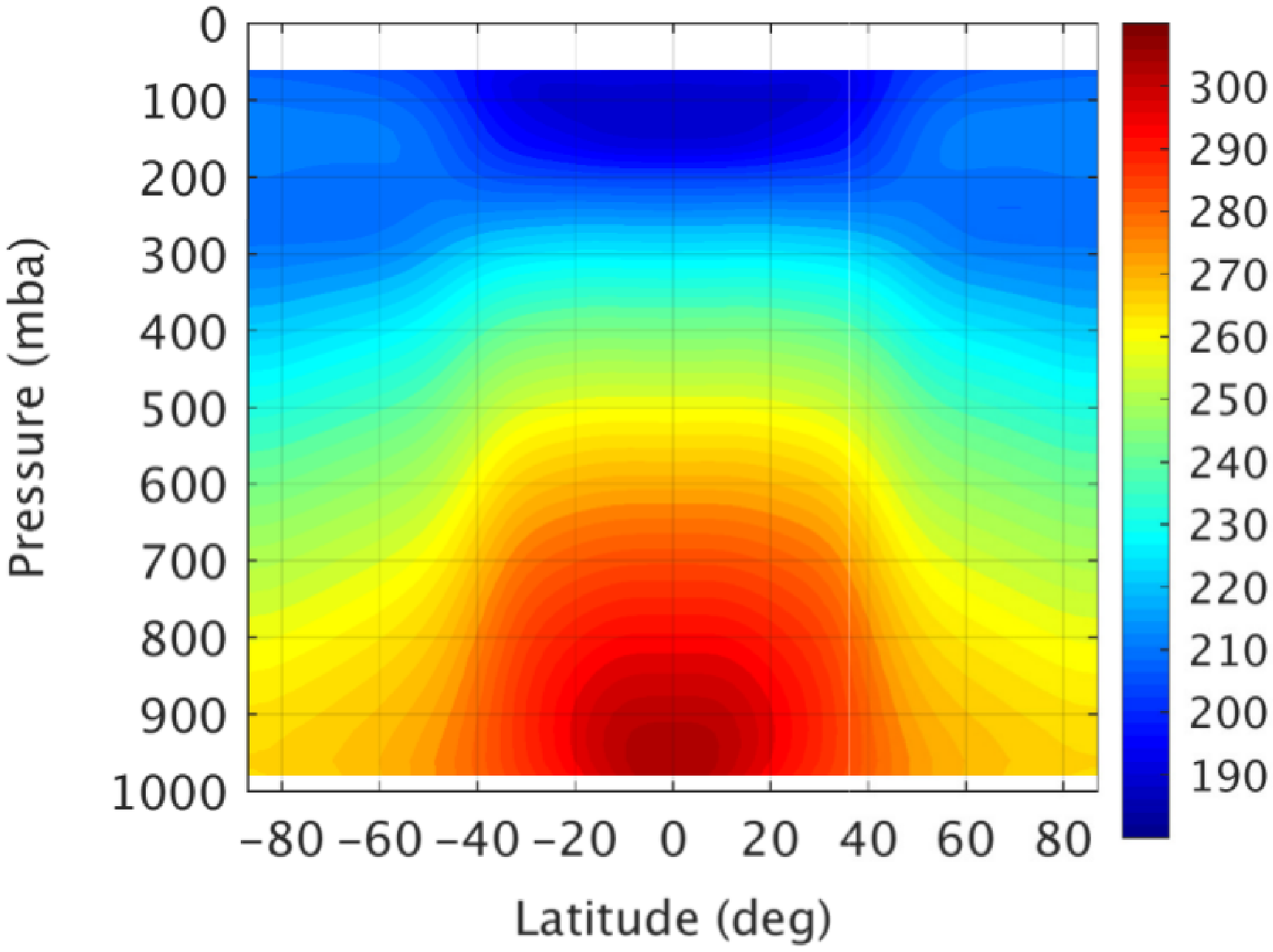}}
\subfigure[Potential temperature]{\label{fig:pt-hs}\includegraphics[width=0.5\textwidth]{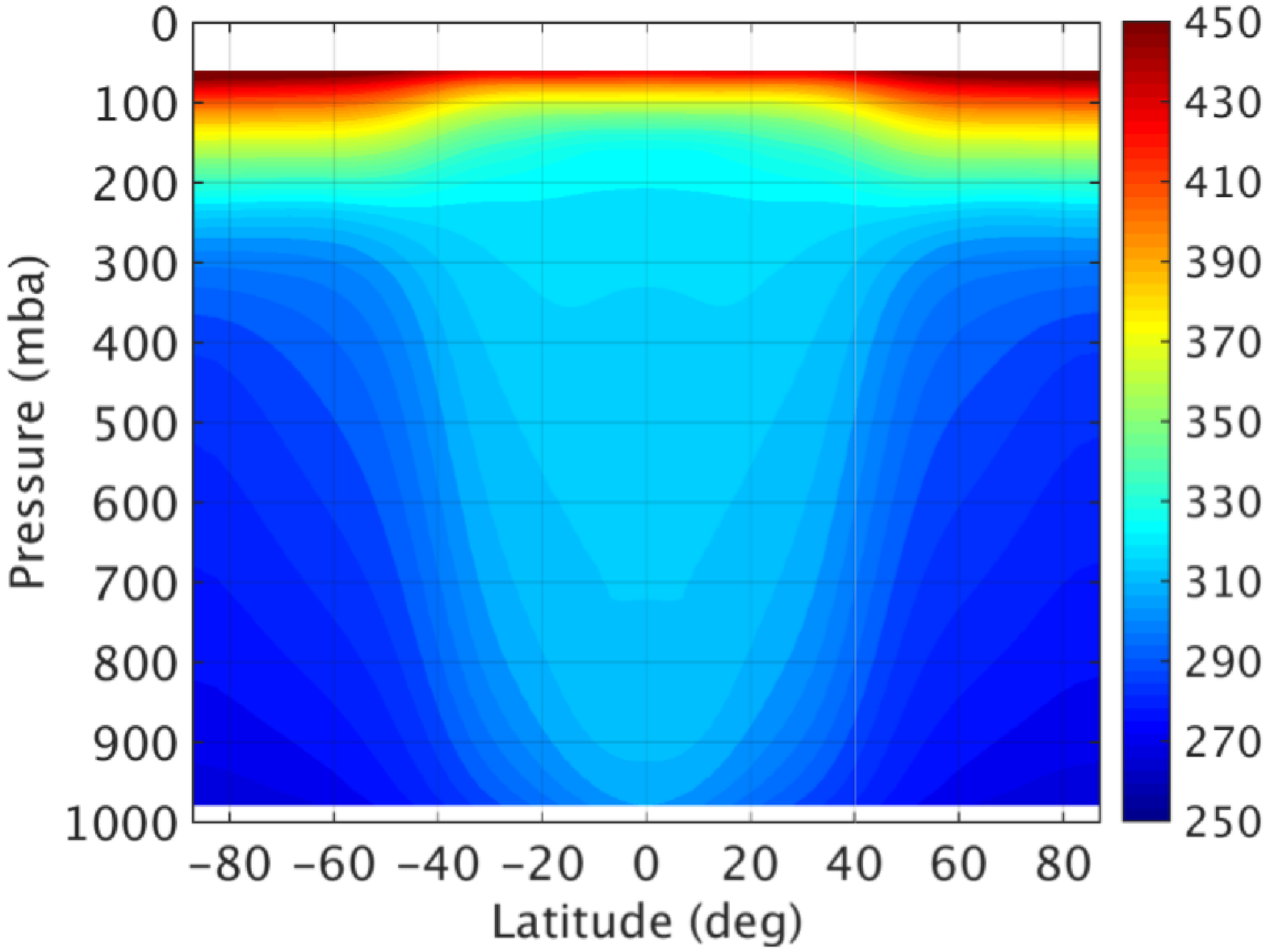}}
\caption{Final averaged temperature and potential temperature (K) for the Held-Suarez experiment. The values were time averaged for 1000 Earth days.}
\label{fig:HS-EXP-TS}
\end{figure}

The averaged temperature and potential temperatures are shown in Fig. \ref{fig:HS-EXP-TS}. These two averaged maps are very similar to the ones shown in \cite{1994Held}. The resulted simulated potential temperature shows weaker vertical gradient at the tropics than at higher latitudes. This configuration indicates weaker static stability at lower latitudes, which drives more upward motion in that region. The cold stratospheric region is obtained due to the constant basic state temperature in the temperature forcing scheme, which makes that region of the atmosphere well stratified and inactive.

\begin{figure}
\centering
\subfigure[Meridional eddy flux of zonal momentum]{\label{fig:me-zm-mr}\includegraphics[width=0.5\textwidth]{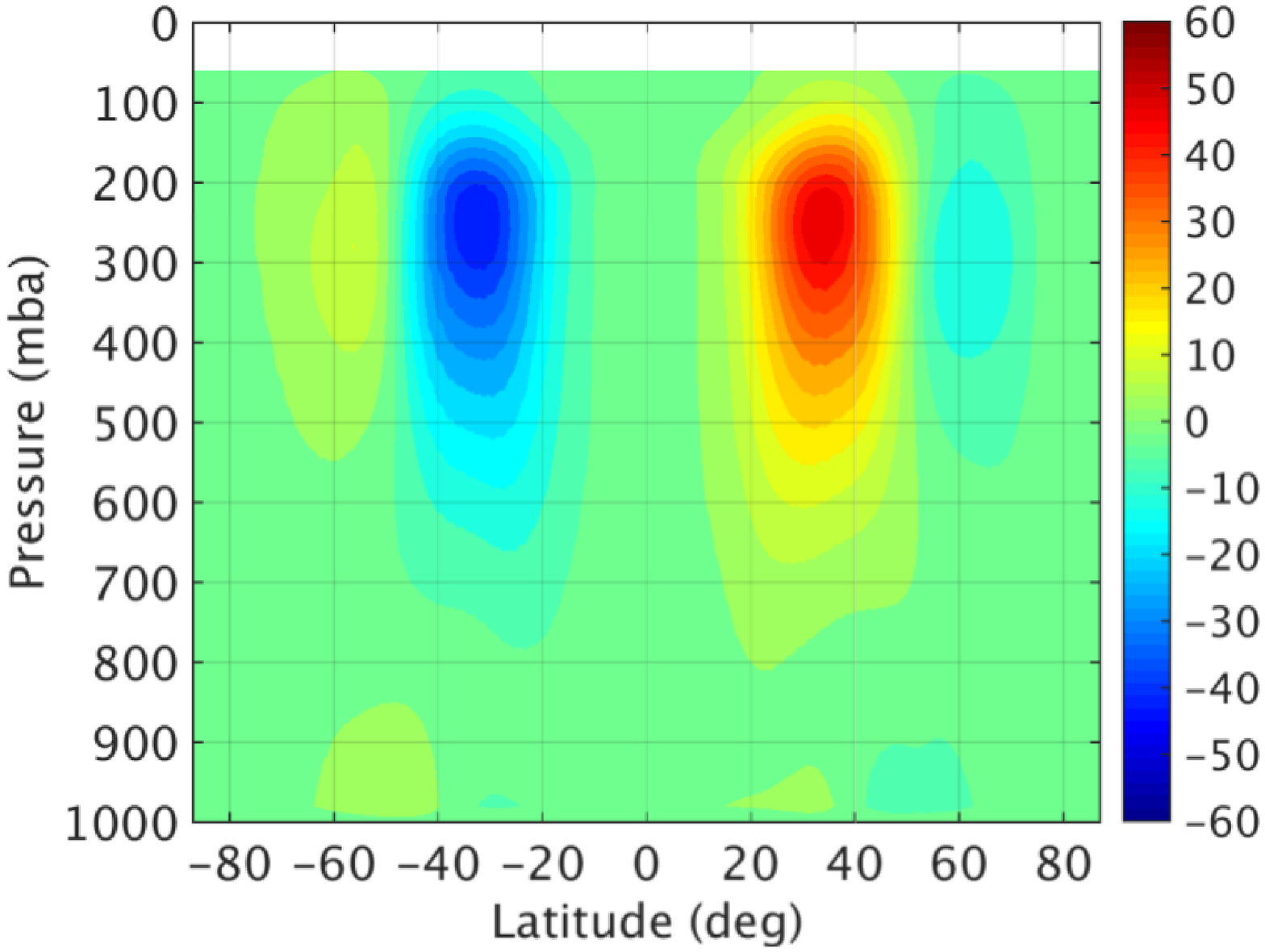}}
\subfigure[Meridional eddy flux of temperature]{\label{fig:me-t-mr}\includegraphics[width=0.5\textwidth]{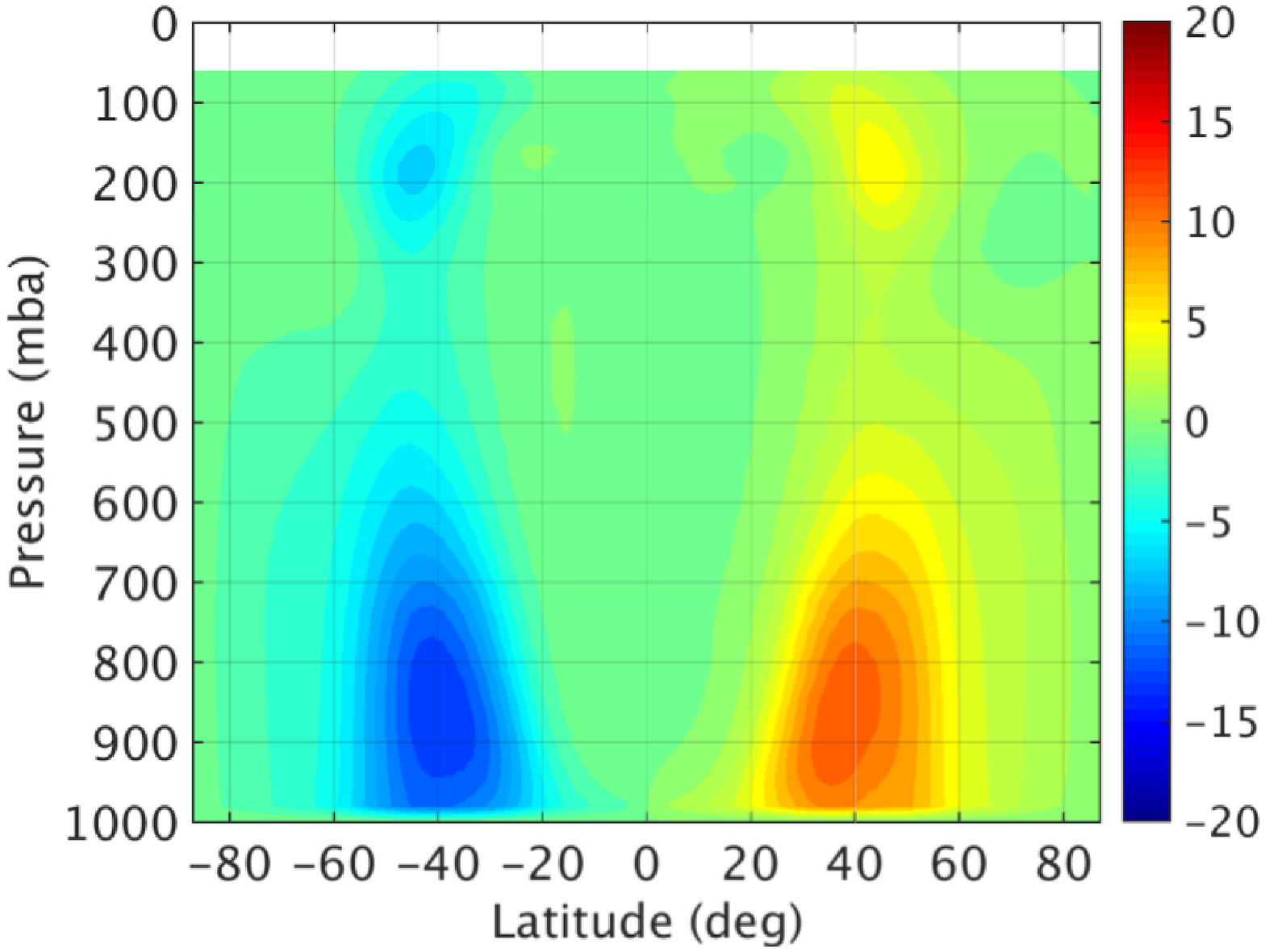}}
\caption{Final averaged meridional eddy flux of zonal momentum ($m^2 s^{-1}$) and temperature (m s$^{-1}$ K) for the Held-Suarez experiment. The values were zonal and time averaged for 1000 Earth days.}
\label{fig:HS-EXP-MEZT}
\end{figure}

\begin{figure}
\centering
\subfigure[Eddy kinetic energy]{\label{fig:edk-hs}\includegraphics[width=0.5\textwidth]{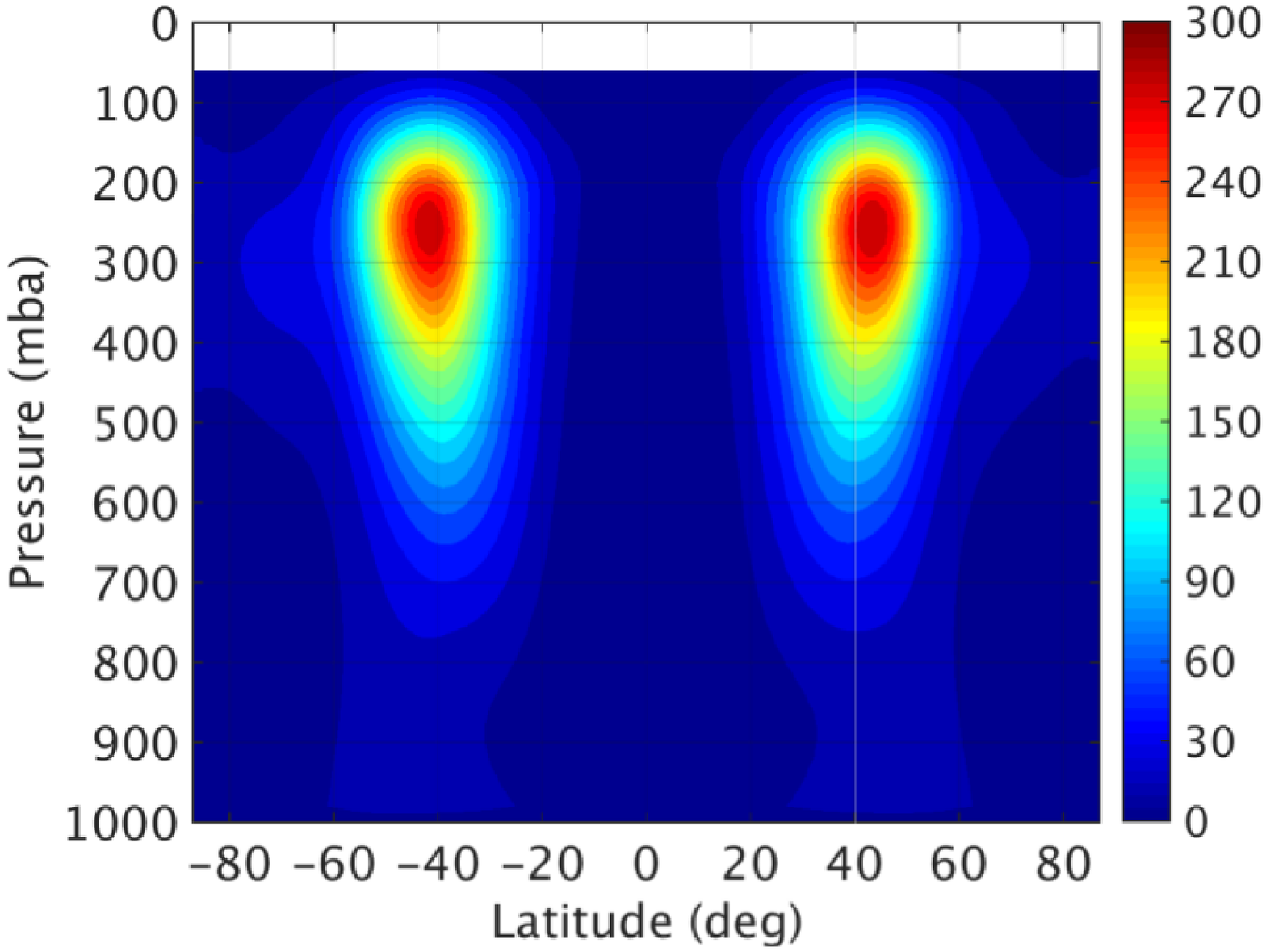}}
\subfigure[Eddy temperature variance]{\label{fig:edt-hs}\includegraphics[width=0.5\textwidth]{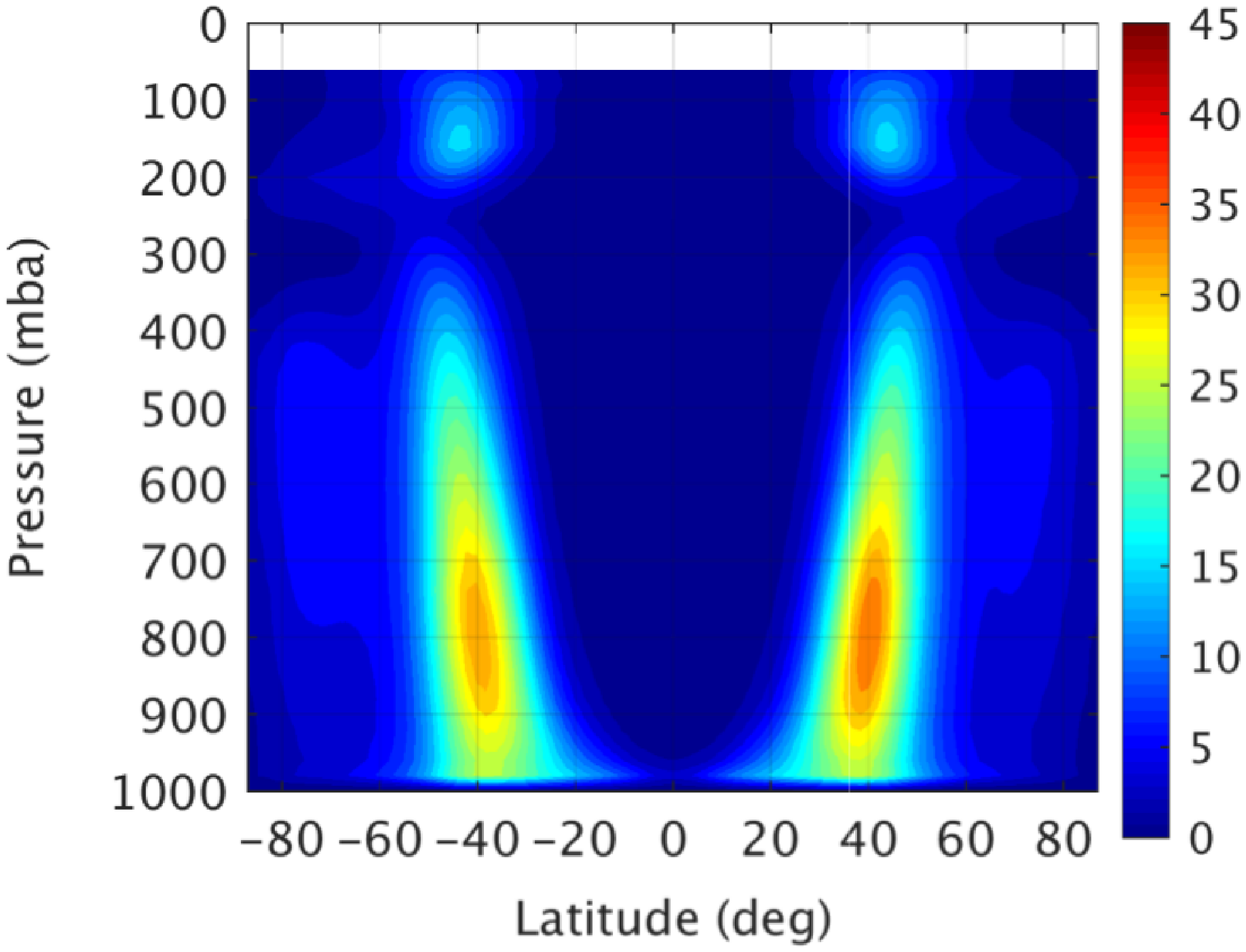}}
\caption{Final averaged eddy kinetic energy ($m^2s^{-2}$) and temperature variance ($K^2$) for the Held-Suarez experiment. The values were time averaged for 1000 Earth days.}
\label{fig:HS-EXP-ETKEV}
\end{figure}

Figs. \ref{fig:HS-EXP-MEZT} and \ref{fig:HS-EXP-ETKEV} are related with wave activity present in the simulation. This activity has a crucial role in the atmospheric circulation produced, and the results are validated comparing to results from \cite{1994Held}, \cite{1998Jablonowski}, and \cite{2012Ullrich}. The zonal inhomogeneities or eddy components of the variables are represented with the dash symbol:
\begin{equation}
 \phi' = \phi - \overline{\phi},
\end{equation}
where $\overline{\phi}$ represents the zonal averaged quantity of $\phi$. Fig. \ref{fig:HS-EXP-MEZT}(a) shows the meridional eddy flux of the zonal momentum. This flux has an important weight in the local balance of the momentum and in the global circulation produced. The flux is more intense at pressure levels above 800 mbar, and in general the zonal motion is transported from low latitudes towards mid latitudes. The divergence of the meridional eddy flux at low latitudes results in a persistent presence of a retrograde flow in that region. In the polar regions, waves transport zonal motions towards low latitudes that converges at mid-latitudes. This convergence combined with the mean circulation form the prograde winds at mid-latitude seen in Fig. \ref{fig:HS-EXP-UV}(a), with one local maximum in each hemisphere at roughly 250 mbar. The eddy heat flux is shown in Fig. \ref{fig:HS-EXP-MEZT}(b). The heat is being transported by atmospheric waves in general from the equatorial regions towards higher latitudes. This is an important quantity for the energy balance of the atmosphere. The heat transport is strongest in the lower atmosphere, where the largest baroclinic wave activity is present.

The eddy kinetic energy and eddy temperature variance quantifies the magnitude of the wave activity in the momentum and temperature fields. As expected from Figs. \ref{fig:HS-EXP-MEZT}(a) and \ref{fig:HS-EXP-MEZT}(b), the largest amplitudes are located at mid-latitudes. In the eddy kinetic energy case, Fig. \ref{fig:HS-EXP-ETKEV}(a), the maxima are located in the zonal wind jet region. In Fig. \ref{fig:HS-EXP-ETKEV}(b), the maxima in the temperature variance correspond to the maxima in the baroclinic wave activity in the lower atmosphere seen in Fig. \ref{fig:HS-EXP-MEZT}(a). 

Comparing our results on the mean circulation and wave activity with the results from \cite{1994Held}, we concluded that both results are quantitatively very similar. The similarity between the two models' results demonstrate that our new model has passed this important benchmark test.  

\begin{figure}
\centering
\subfigure[Vertical velocity difference]{\label{fig:w-hs}\includegraphics[width=0.5\textwidth]{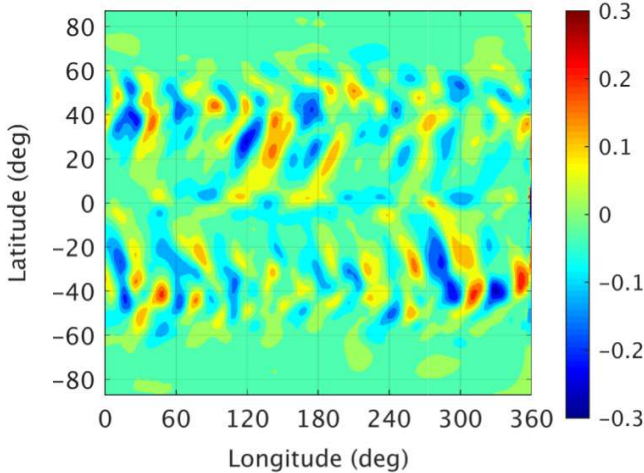}}
\caption{Final time averaged vertical velocity difference ($10^4$m/s) in the first model layer for the Held-Suarez experiment. The values were time averaged for 1000 Earth days.}
\label{fig:HS-EXP-VD}
\end{figure}

The geometrical grid properties can be associated with sources of numerical noise visible in the solution (see section  \ref{subsec:modicogr}). In this work, this unphysical quantity, called grid imprinting, is reduced from the numerical solution when we applied the methods that smoothed the distortions in the standard icosahedral grid (``spring dynamics method'') and moved the points to the centroids of the control volumes. To verify the impact of the grid imprinting in our simulation, we analyzed a map of the vertical velocity difference ($[w^*]$) averaged over the last 1000 Earth days of the long simulation (Figure \ref{fig:HS-EXP-VD}). $w^*$ is defined as,
\begin{equation}
w^* = w - [w].   
\end{equation}
The square brackets represent the time averaging operator. This quantity in the first layer is a good diagnostic for grid imprinting because it is much smaller than the other prognostic variables, and in the case of the presence of persistent small perturbations from the grid structure, they would be easily seen from $[w^*]$ results (\citealt{2012Ullrich}). From Fig. \ref{fig:HS-EXP-VD} we do not see any artificial values from the grid structure, which is an indication that the grid imprinting noise is efficiently reduced in our simulations.

\subsection{Benchmark test for hot Jupiters}
\label{sec:HJ}

The new model is aimed at probing a large range of planetary conditions. Here we explore a simulation under the conditions of a typically tidally locked hot Jupiter proposed by \cite{2009Menou}. This experiment was suggested in \cite{2011Heng} as a benchmark test and it was also explored later in \cite{2013Bending} and \cite{2014Mayne}. 

This experiment was integrated without boundary layer friction at the bottom as suggested in \citealt{2009Menou}, which means that the function $K_v$ is set to zero (no interchanges of momentum between the simulated atmosphere and the interior). The temperature equilibrium function ($T_{eq}$) used in Eq. (\ref{eq:im_eq_t}) is defined for this experiment as:

\begin{equation}
\label{eq:teq-hj}
T_{eq} = T_{vert} + \beta_{trop}\Delta T_{E-P}\cos(\lambda)\cos(\phi),
\end{equation}
where $\Delta T_{E-P} = 300$K (equator-to-pole difference), $\lambda$ is the longitude, $\phi$ is the latitude, $\beta_{trop}$ defined by:

\begin{equation}
\label{eq:btrop}
\beta_{trop} = \begin{cases} \sin \frac{\pi(\sigma - \sigma_{stra})}{2(1-\sigma_{stra})}&,  z \leq z_{stra} \\ 0&,  z > z_{stra} \end{cases}.
\end{equation}

In Eq. (\ref{eq:btrop}), $\sigma_{stra} = 0.12$ and $z_{stra} = 2\times 10^3$ km. The function $T_{vert}$ used in Eq. (\ref{eq:teq-hj}) has the following form:

\begin{equation}
\label{eq:tvert}
T_{vert} = \begin{cases} T_{surf} - \Gamma_{trop}\Big(z_{stra} + \frac{z-z_{stra}}{2}\Big)\\
+\Big(\Big[\frac{\Gamma_{trop}(z-z_{stra})}{2}\Big]^2 + \Delta T^2_{stra}\Big)^{1/2}&,  z \leq z_{stra} \\ 
T_{surf}-\Gamma_{trop}z_{stra} + \Delta T_{stra}&,  z > z_{stra} \end{cases},
\end{equation}
where $T_{surf} =$ 1600 K, $\Gamma_{trop}= 2\times10^{-4}$ K/m and $\Delta T_{stra} = 10$ K. The timescale in the temperature forcing scheme is set to a constant value equal to $1.5 \times 10^5$ s. The initial conditions in this experiment were an isothermal atmosphere at 1600 K in hydrostatic equilibrium and the wind velocity components set to zero. The time step used was 600 seconds, a grid refinement equals to g-level 5, 37 vertical layers covering from 0 to 4875 km altitude and a diffusion timescale of 941.2 seconds.

\begin{figure}
\centering
\subfigure[Eastward wind speed]{\label{fig:u-mr}\includegraphics[width=0.5\textwidth]{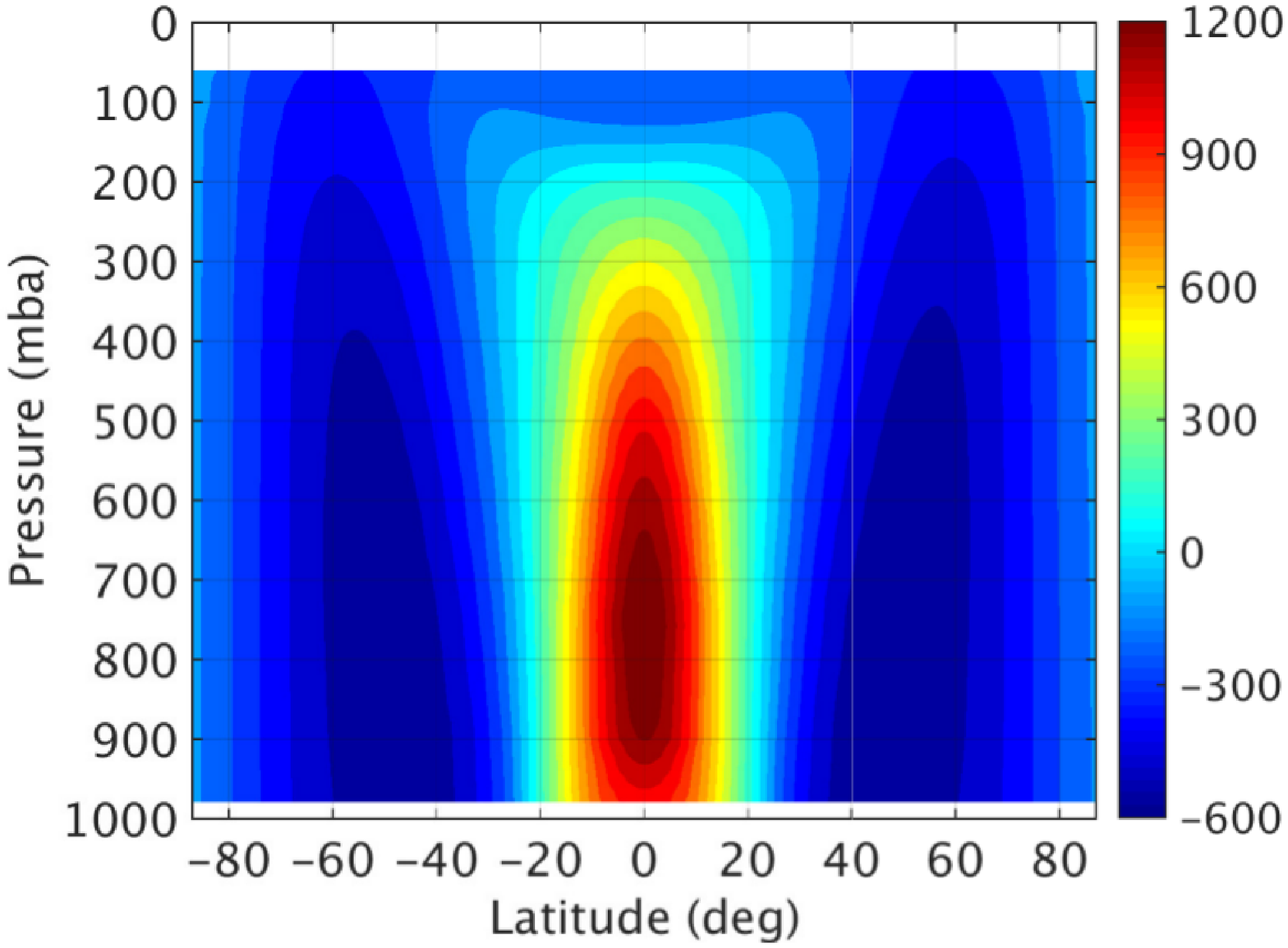}}
\subfigure[Mass stream function]{\label{fig:v-mr}\includegraphics[width=0.5\textwidth]{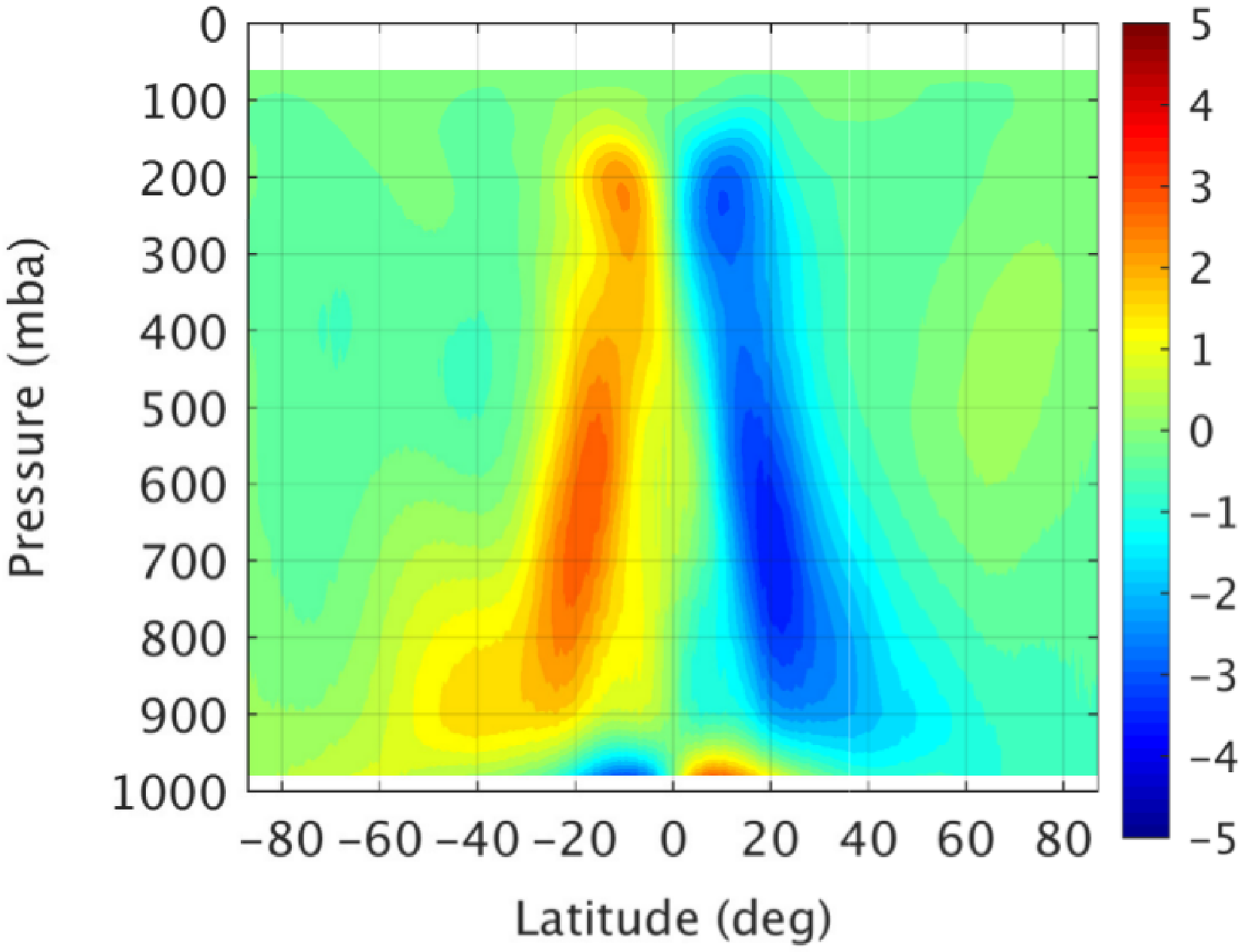}}
\caption{Final averaged zonal winds (m/s) and mass stream function ($10^{13}$ kg/s) for the hot Jupiter experiment. The values were time averaged for 1000 Earth days.}
\label{fig:MR-EXP-UV}
\end{figure}

The atmospheric circulation produced is very distinct from the previous Earth-like case. Here the atmospheric circulation is largely influenced by the strong day-night contrast in the temperature forcing. Fig. \ref{fig:MR-EXP-UV}(a) shows the zonal winds averaged in longitude and in time over the last 1000 Earth days of the simulation. In general, the zonal winds show a strong prograde jet at low latitudes and centered at a pressure level of 800 mbar.  The mass stream function shows the presence of indirect atmospheric cells (eddy-driven cells) at low latitudes with two peaks in the mass stream function in each hemisphere at different pressure levels: roughly 800 mbar and 250 mbar. This mean circulation combined with the excited equatorial waves transport prograde axial angular momentum towards low latitudes and help maintaining the strong prograde equatorial jet. At higher latitudes the direction of the winds reverse. These retrograde winds are expected from the conservation of the total axial angular momentum in the atmosphere. In these regions, the mass streamfunction map shows that the flow follows on average a clockwise direction in the northern hemisphere and anticlockwise in the southern hemisphere. 

\begin{figure}
\centering
\subfigure[Temperature]{\label{fig:t-mr}\includegraphics[width=0.5\textwidth]{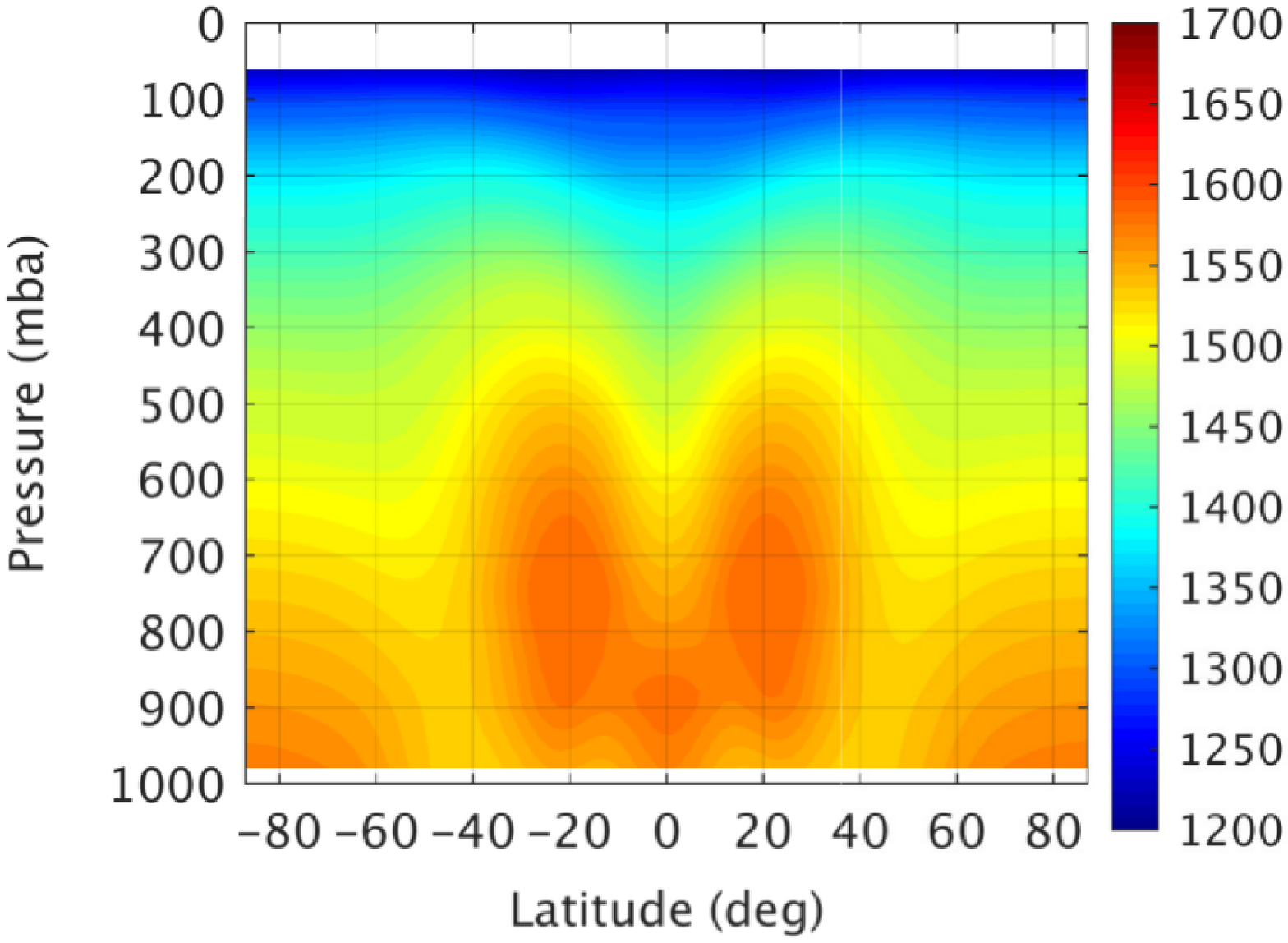}}
\subfigure[Temperature at sigma level 0.7]{\label{fig:pt-mr}\includegraphics[width=0.5\textwidth]{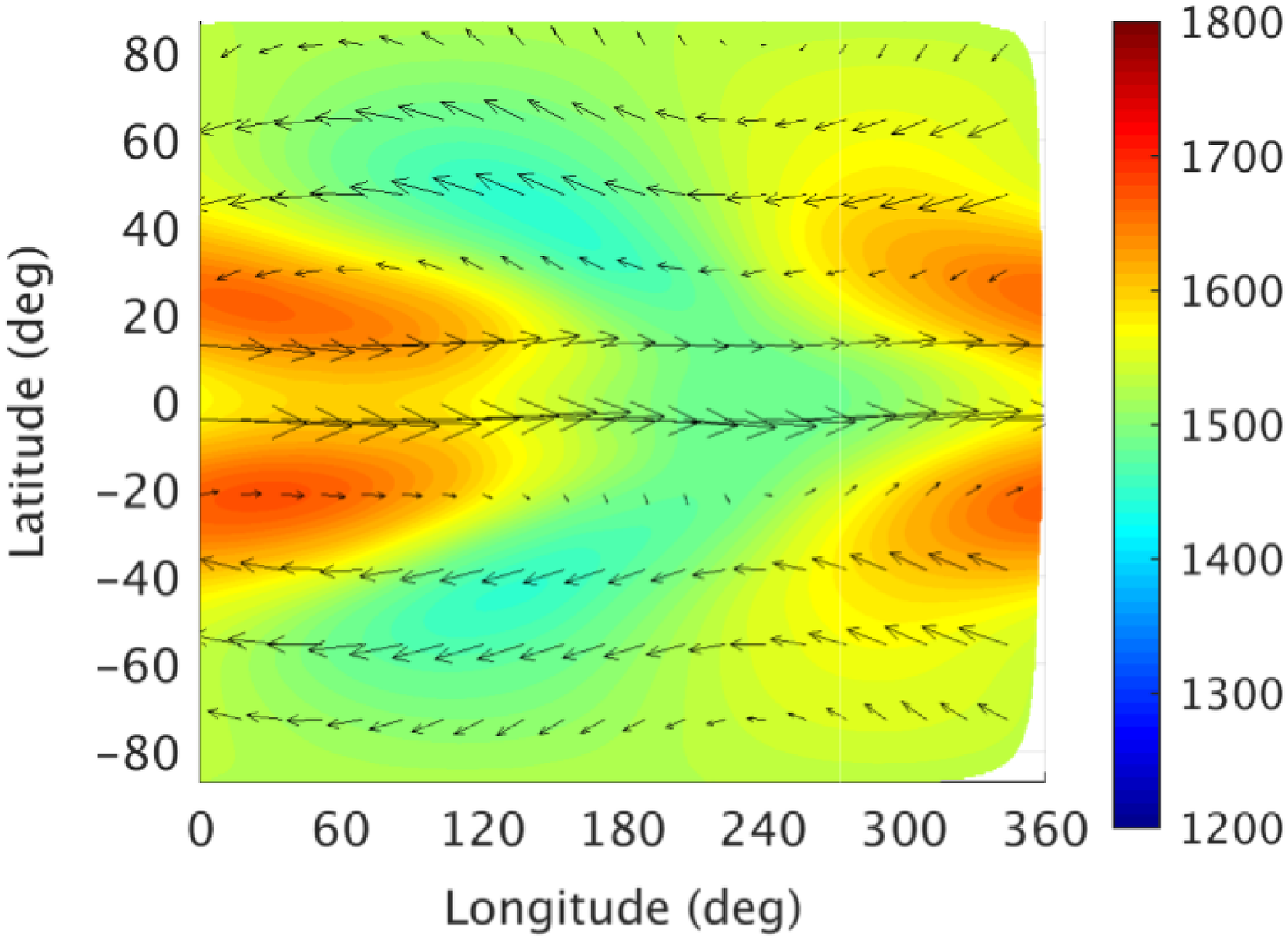}}
\caption{\textbf{(a)} Final averaged temperature (K) for the hot Jupiter experiment. \textbf{(b)} Final time averaged horizontal map of temperature at sigma level 0.7. The arrows shows the time averaged direction of the wind speed. In both maps the values were averaged for 1000 Earth days.}
\label{fig:MR-EXP-TSAV}
\end{figure}

\begin{figure}
\label{fig:temp-lev-shj}
\centering
\subfigure[Day 346]{\includegraphics[width=0.437\textwidth]{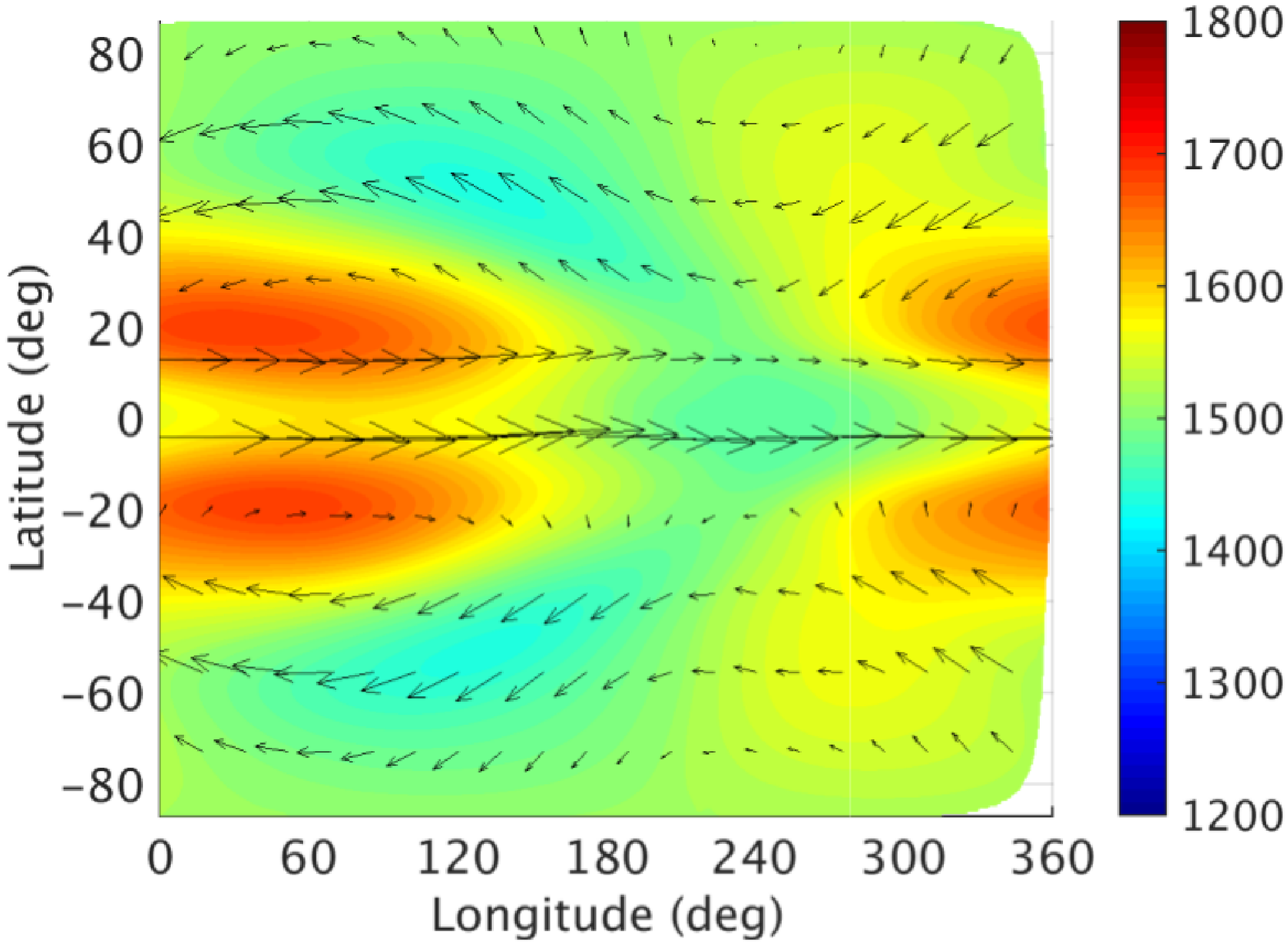}}
\subfigure[Day 353]{\includegraphics[width=0.437\textwidth]{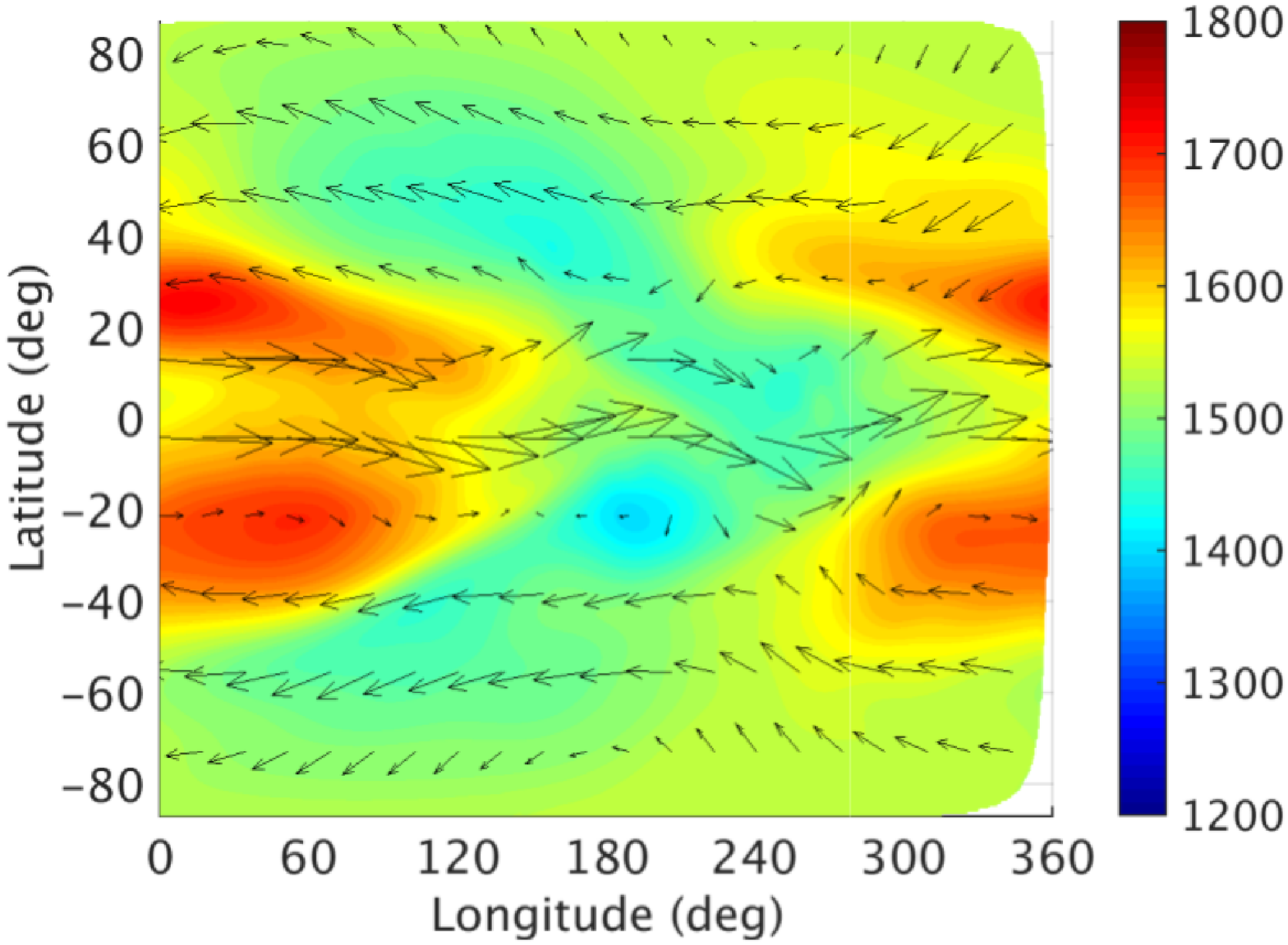}}
\subfigure[Day 360]{\includegraphics[width=0.437\textwidth]{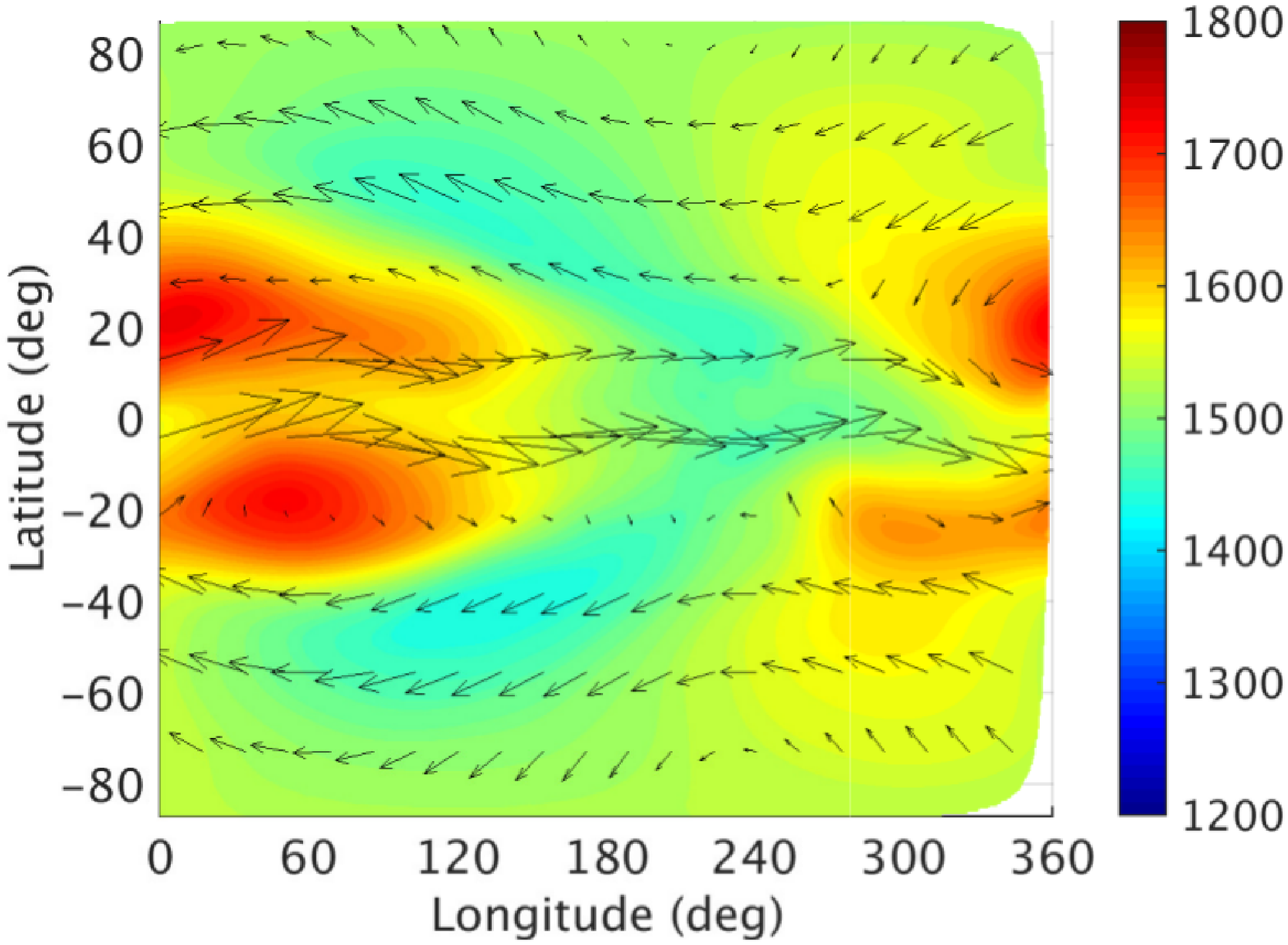}}
\caption{Snapshots of the temperature field at sigma 0.7 for the days 346, 353 and 360. \cite{2009Menou} and \cite{2011Heng} only show day 346. The arrows show the direction of the circulation.}
\label{fig:MR-EXP-TS}
\end{figure}

The averaged temperature map is shown in Fig. \ref{fig:MR-EXP-TSAV}(a). In this map, we see clearly two maxima in the temperature in each hemisphere, which are related with the efficient heat transport at the equatorial region. Fig. \ref{fig:MR-EXP-TSAV}(b) shows a time averaged map of the temperature and winds at the pressure level 700 mbar. The two maxima in the temperature in each hemisphere are clear in this map as well. The heat is being transported in opposite directions at low (prograde transport) and high latitudes (retrograde transport). At low latitudes, it is efficiently transported by the broad strong equatorial jet. Fig. \ref{fig:MR-EXP-TS} shows snapshots of the temperatures and winds at sigma level 0.7.  The map at day 246 has been used in \cite{2011Heng}, \cite{2013Bending} and \cite{2014Mayne} for comparison with the work of \cite{2009Menou}. However, one should be careful when comparing instantaneous results due to the variability of the flow.

In general, the results obtained with our model are quantitatively very similar to the ones obtained by \cite{2009Menou}, \cite{2011Heng} and \cite{2013Bending}. Note that the sub-stellar point in our model is shifted by $\pi$ when compared with \cite{2009Menou} and \cite{2011Heng} results. The results obtained in \cite{2014Mayne} are qualitatively similar to the other models, but quantitatively very different. The main differences are at the tropics. The magnitude of the equatorial jet is $40\%$ weaker and is located much lower in pressure (~400 mbar). These differences in the results have an impact on the heat transport of the simulated atmosphere, as it is clear when comparing their averaged temperature field results with the other models (including our model). In \cite{2014Mayne}, the differences in the model domain and boundary conditions applied are suggested as possible causes for the differences in the results. The differences in the domain are attributed to the use of  different vertical coordinates between the models: pressure coordinates in \cite{2009Menou} and \cite{2011Heng}, and height in \cite{2014Mayne}. However, in our model we use the same model domain and boundaries as in \cite{2014Mayne} and our results are very similar to \cite{2009Menou} and \cite{2011Heng} results. For this reason, the cause for the disagreement in the results between \cite{2014Mayne} and the other models remain unclear.

\newpage
\section{Conclusion}
\label{sec:conclusion}
We have developed a new global circulation model for a large diversity of atmospheric conditions. This new platform has passed two important benchmark tests with very different atmospheric conditions: for Earth and for a hot Jupiter-like planet. The grid implemented is able to reach a second-order accuracy and the grid imprinting noise is efficiently removed from the simulations.  In this, work we have described the structure of the new complex dynamical model that uses a numerical solver based on an explicit-implicit formulation. This formulation allows for a very good performance since we use a split-explicit time step method coupled with an implicit integration of the vertical momentum, which is numerically constrained by the propagation of the fast sound waves across the fine vertical resolution. The model does not use any of the traditional approximations often used in Earth climate studies that can limit the model flexibility to explore other planetary conditions. This new platform solves the deep non-hydrostatic Euler equations, and more physical representations can be easily added to the original set of equations.

The two experiments which were successfully done represent a small sample in the general diversity of planets. However, \texttt{THOR} showed a very good performance in the simulation of these two very distinct atmospheric conditions, which allow us to be in an advantageous position to continue exploring the diversity of planetary parameters that are expected to characterize the atmospheres of solar system and extrasolar planets. 

In summary, the main advantages of using our new platform against other recent planetary models are:
\begin{itemize}
\item The resolved atmospheric fluid flow is completely represented and no approximations are used that could compromise the physics of the problem; 
\item The model uses, for the first time in exoplanetary studies, an icosahedral grid that solves the pole problem; 
\item The interface is user friendly and can be easily adapted to a multitude of atmospheric conditions. 
\end{itemize}

We have developed a solid basis for a virtual atmospheric laboratory. This model has also been developed keeping the user interface easy to use since we aim to make this code free and open-source to the community (more details about \texttt{THOR} in www.exoclime.net). 
\newpage
%  ACKNOWLEDGMENTS
\section*{Acknowledgments}
J.M.M., S.G., L.G. and K.H. acknowledge financial and administrative support from the Center for Space and Habitability
(CSH), the PlanetS National Center of Competence in Research (NCCR), the Swiss National Science Foundation (SNF) and the Swiss-based MERAC Foundation. \texttt{THOR} is part of the Exoclimes Simulation Platform (ESP) that will be freely available to the community (see www.exoclimes.net).

%%% REFERENCES %%%
\bibliography{mybibfile}

\end{document}